\documentclass[acmsmall]{acmart}
\AtBeginDocument{%
  }

\setcopyright{cc}
\setcctype{by}
\copyrightyear{2026}
\acmYear{2026}
\acmDOI{10.1145/3857797}

\acmJournal{CSUR}
\acmVolume{1}
\acmNumber{1}
\acmArticle{}
\acmMonth{1}
\usepackage{multirow}
\usepackage{subcaption}
\usepackage{listings}
\usepackage{wrapfig}

\usepackage{lscape}
\usepackage{makecell}

\begin{document}
\title{FPGA-Based Neural Network Accelerators for Space Missions: A Survey}

\author{Pedro Antunes}
\correspondingauthor
\email{pedroa@kth.se}
\orcid{0000-0002-6158-6928}
\author{Artur Podobas}
\orcid{0000-0001-5452-6794}
\email{podobas@kth.se}
\affiliation{%
  \institution{KTH Royal Institute of Technology}
  \department[1]{School of Electrical Engineering and Computer Science}
  \department[0]{Department of Computing and Learning Systems}
  \city{Stockholm}
  \country{Sweden}
}

\renewcommand{\shortauthors}{Antunes and Podobas}

\begin{abstract}
  Space missions are growing ever more ambitious, placing greater demands on onboard computing. Field-programmable gate arrays (FPGAs) have garnered interest due to their reconfigurability and cost-effectiveness. At the same time, neural network (NN)-based methods are proving invaluable for critical spacecraft tasks such as autonomous operations, remote sensing, selective downlink, and data compression. This survey reviews and classifies the state of the art in FPGA-based NN accelerators for space missions. We examine current trends, highlight key challenges, and suggest directions for future research. This work highlights key aspects for integrating high-performance NN acceleration into next-generation spacecraft computing systems.
\end{abstract}



\received{May 30 2025}
\received[revised]{Sep 7 2026}
\received[accepted]{Sep 28 2026}

\maketitle

\section{Introduction}
\label{sec:Introduction}
Space -- the final frontier -- has captivated human curiosity for centuries. The dawn of the space race marked a turning point, leading to the exploration of nearly every major celestial body in our solar system: Mars (Viking 1/2), Venus (Venera 7), Mercury (Mariner 10), Jupiter (Galileo), Saturn (Cassini), and various moons (e.g., Apollo 11, Phobos 2). This momentum continues today with numerous ongoing and planned missions.\footnote{\url{https://www.astronomy.com/space-exploration/space-missions-a-list-of-current-and-upcoming-voyages/}} Simultaneously, falling launch costs~\cite{jonesRecentLargeReduction2018} have accelerated the deployment of small satellites, including nanosatellites and CubeSats—a trend showing no signs of slowing~\cite{kuluSmallLaunchers20232023}.

Onboard computer systems are at the heart of every spacecraft, from CubeSats to interplanetary probes. These systems handle everything from navigation and control -- such as the Apollo Guidance Computer~\cite{o2010apollo} -- to the analysis of data from increasingly sophisticated sensors -- as in European Space Agency (ESA)'s $\phi$-Sat-1 mission~\cite{giuffridaFSat1MissionFirst2022}. As missions grow more complex and data-intensive, the need for high-performance onboard computing becomes increasingly critical. However, the deployment of computing systems in space missions is inherently stricter than in other fields~\cite{space_computing,space_fault_tolerant_survey}. Spacecraft operate in a harsh environment~\cite{jamesNaturalSpaceEnvironment1994} characterized by high levels of ionizing radiation~\cite{edmondsIntroductionSpaceRadiation2000,SEE_semiconductor_survey}, intense mechanical vibrations during launch~\cite{vibrations_lunch}, extreme temperature fluctuations~\cite{gpu_heat_space}, and a stringent lack of accessible energy sources~\cite{energy_space}. These constraints make reliable and power-efficient execution of demanding tasks, such as deep learning inference, exceptionally challenging and necessitate specialized hardware setups~\cite{space_computing}.

Consequently, field-programmable gate arrays (FPGAs) now play a central role in spacecraft computing~\cite{kuon2008fpga}. Their reconfigurability allows the hardware design to be developed after chip fabrication and later mapped onto the device. This flexibility allows FPGAs to instantiate customized hardware architectures tailored to neural network (NN)~\cite{dnn_basics} workloads. By deploying these purpose-built accelerators near the sensors, spacecraft can process incoming data in real time, maximizing computational performance per watt and establishing FPGAs as a platform for onboard artificial intelligence (AI). Radiation-tolerant device families~\cite{PolarFire,xilinx2020rt} further enhance the attractiveness of FPGAs under spacecraft constraints.

Researchers envision AI in space since the Deep Space 1 mission~\cite{hedbergAIComingAge1997}. Since then, advances in deep learning sparked interest in onboard NN deployment, with ESA's $\phi$-Sat-1 as the first mission to integrate this concept. AI is essential for applications such as autonomous operations, remote sensing~\cite{zhuDeepLearningRemote2017}, data compression~\cite{data_compression}, and selective downlinking~\cite{rapuanoFPGABasedHardwareAccelerator2021}. These applications are vital for reducing latency and managing bandwidth in data-constrained missions.

This survey reviews the current state of FPGA-based NN accelerators for space missions. While prior surveys have covered NNs in general~\cite{wangConvergenceEdgeComputing2020,huangDeepLearningBasedSemanticSegmentation2024}, FPGA-based CNN accelerators~\cite{mittalSurveyFPGAbasedAccelerators2020}, AI in space missions~\cite{ocheApplicationsChallengesArtificial2024,chienFutureAISpace2006}, and recently FPGA-based Machine Learning (ML) implementations for remote sensing applications~\cite{Leonard_2026}, to our knowledge, our work is the first comprehensive review explicitly focused on FPGA-based NN accelerators for space. In light of this, the main contributions of this survey are:
\begin{itemize}
    \item A comprehensive classification and review of the state-of-the-art FPGA-based NN accelerators proposed for space applications.
    \item An analysis of the methods (e.g., quantization, pruning, and fault-tolerance mechanisms) used to meet the power, performance, and reliability constraints of the space environment.
    \item A critical discussion of current trends, highlighting the most pressing technical challenges and identifying promising directions for future research in onboard AI acceleration.
\end{itemize}

The remainder of this paper is structured as follows: Section~\ref{sec:Section_2} presents background material; Section~\ref{sec:Section_3} outlines our survey methodology and summarizes the literature; Section~\ref{sec:Section_4} discusses key insights, trends, and research gaps; and Section~\ref{sec:Section_5} concludes the paper.

\section{Background}
\label{sec:Section_2}
Space missions span near-Earth orbits, such as Low Earth Orbit (LEO), and deep space beyond the Moon~\cite{barthSpaceAtmosphericEnvironments2003,MEO,unitedstatesTitle42United2023}. These environments differ in radiation exposure, which can degrade computing hardware~\cite{jamesNaturalSpaceEnvironment1994,edmondsIntroductionSpaceRadiation2000}. Launch vibrations, thermal cycling, and limited energy also shape spacecraft design~\cite{vibrations_lunch,energy_space}. However, in this survey, we focus on hardware-logic implications.

\subsection{Hardware for Space Missions}
Satellite hardware integrates CPUs, GPUs, FPGAs, and Application-Specific Integrated Circuits (ASICs). CPUs offer high flexibility but often lack the performance required for compute-intensive tasks. GPUs enable high parallelism but typically consume significantly more power. In space, this elevated power consumption translates into a thermal dissipation bottleneck~\cite{gpu_heat_space} and often makes them unsuitable for tight thermal and power budgets. By contrast, ASICs offer the highest power efficiency, minimizing both energy consumption and thermal dissipation, but they entail high manufacturing costs and long development cycles. Moreover, their production volume may be economically infeasible, and once launched, their functionality cannot be modified. FPGAs balance customization and cost-effectiveness, making them attractive for many space missions.

While commercial off-the-shelf (COTS) versions of these components are sometimes used for short-lived space missions, they are typically inadequate for harsh radiation environments~\cite{COTS_in_Space}. These environments expose hardware to radiation that can cause short-term effects known as Single Event Effects (SEEs) and long-term degradation. The latter is primarily associated with the device's Total Ionizing Dose (TID) response. TID sensitivity mainly depends on the underlying semiconductor technology rather than the hardware architecture or the implemented algorithm~\cite{TID_CMOS_space}. SEEs often manifest as bit flips in memory or logic, which can propagate through the system and lead to incorrect behavior~\cite{SEE_semiconductor_survey}. To mitigate these radiation-induced effects, systems can be designed with fault-tolerant components or dedicated radiation-tolerant devices, such as the Kintex UltraScale~\cite{xilinx2020rt} and PolarFire~\cite{PolarFire}. Although these devices are more costly and may offer lower raw performance than COTS components, they enhance reliability in space applications~\cite{space_computing,space_fault_tolerant_survey}.

\subsection{Reconfigurable Hardware}
\begin{wrapfigure}{r}{0.5\textwidth}
  \centering
  \includegraphics[width=\linewidth]{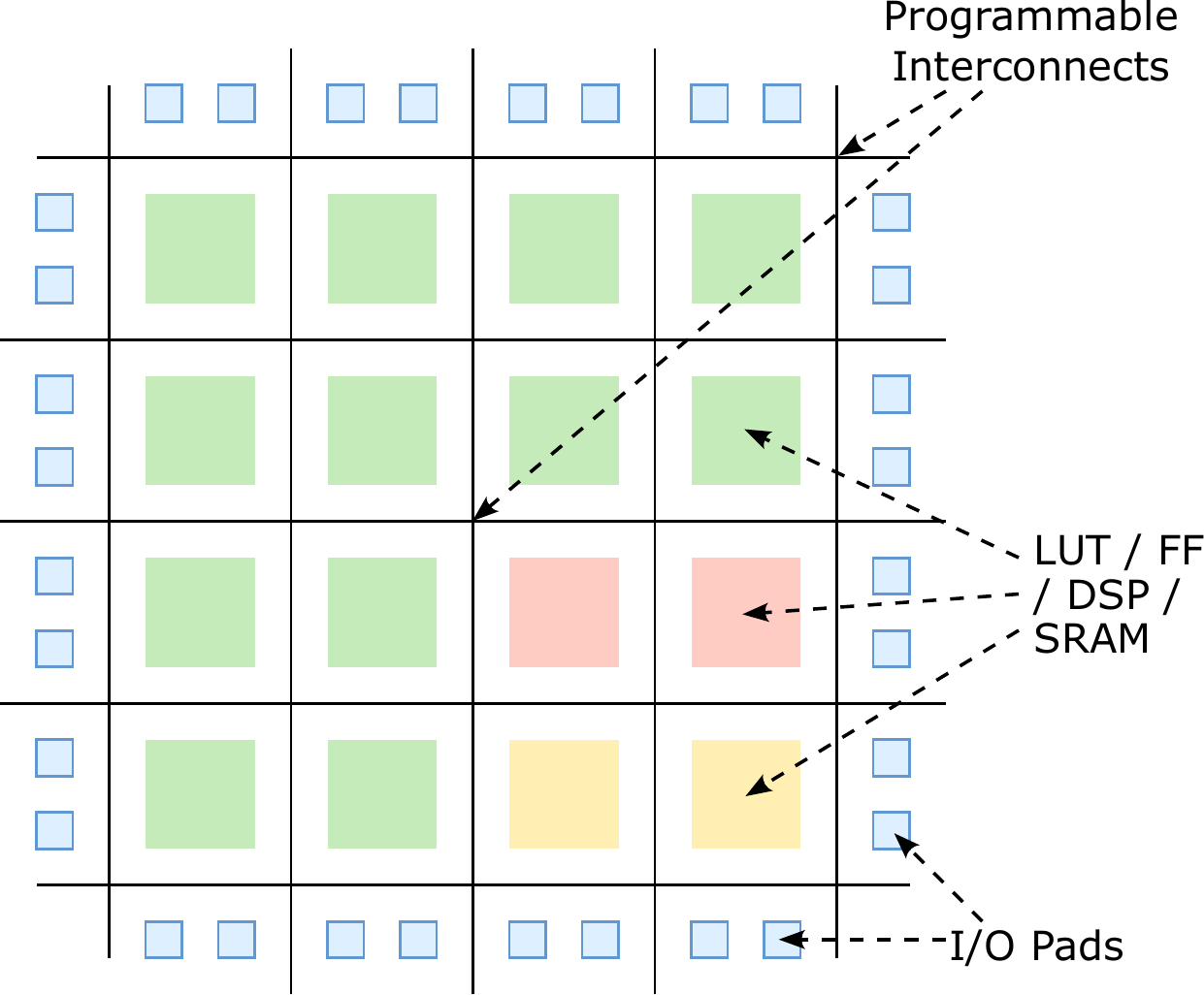}
  \caption{\label{fig:fpga_layout}Abstract representation of an FPGA, showing logic blocks, programmable interconnect, and IOs}
  \Description[Abstract representation of the structure of an FPGA, showing logic blocks, programmable interconnect, and IO pad.]{Abstract representation of the structure of an FPGA, showing logic blocks, programmable interconnect, and IO pad. The figure illustrates the internal architecture of an FPGA, highlighting its key components. These components include logic blocks, the basic building blocks of the FPGA; programmable interconnects that enable flexible connections between logic blocks; and input/output (I/O) pads that facilitate communication with external devices.}
\end{wrapfigure}
Reconfigurable hardware preserves a degree of flexibility typically sacrificed in ASICs. This reconfigurability manifests at different levels of granularity~\cite{podobas2020survey}, with FPGAs being the most prevalent and fine-grained option. FPGAs feature thousands or millions of look-up tables (LUTs), programmable interconnects, and Input/Output (I/O) pads (Figure~\ref{fig:fpga_layout})~\cite{FPGAs_architecture}. Programming these LUTs with Boolean functions and configuring the interconnects maps a digital circuit onto the fabric without fabricating new silicon. While early FPGAs contained only LUTs, modern devices add dedicated DSP blocks for arithmetic and on-chip static random-access memory (SRAM) to improve performance~\cite{FPGAs_architecture}. These devices are employed across diverse computing domains, including ASIC prototyping, telecommunications, low-volume consumer electronics, high-performance computing, and space applications~\cite{fpga_uses}. Some systems combine FPGAs with hard processor cores, resulting in System-on-Chip (SoC) configurations that encompass programmable logic (PL) and a processing system (PS). Compared to ASICs, FPGAs can offer cost-effectiveness, radiation tolerance, remote reconfiguration, and reduced time-to-market \cite{boadagardenyesTrendsPatternsASIC2011}.

Two primary methodologies exist for designing systems on FPGAs. The first involves using hardware description languages (HDLs), such as VHDL, Verilog, or Chisel, to manually detail designs at the Register Transfer Level (RTL). This approach offers precise control over design descriptions but comes with a steep learning curve. Alternatively, systems can be described in abstract programming languages like C/C++ to define functionality. This method employs High-Level Synthesis (HLS) tools, such as Vitis HLS~\footnote{\url{https://www.amd.com/en/products/software/adaptive-socs-and-fpgas/vitis/vitis-hls.html}} or Intel OpenCL~\cite{czajkowski2012opencl}, which transform abstract C code into RTL. Although HLS tools offer a less daunting learning curve, they provide reduced control over the generated hardware because many decisions are delegated to the HLS compiler. Recently, HLS tools targeting NNs have emerged, for example, FINN~\cite{umurogluFINNFrameworkFast2017} and HLS4ML~\cite{hls4ml}. These tools convert NN models into RTL.

\subsection{Artificial Intelligence in Space}
Integrating AI in space missions enables autonomy, adaptability, and efficiency in environments where human intervention is impractical. Existing surveys extensively explore the broad applications of AI in space missions~\cite{ocheApplicationsChallengesArtificial2024,chienFutureAISpace2006}. Below, we summarize the role of AI in three key domains:
\begin{enumerate}
  \item \textbf{Spacecraft Autonomy:} AI-driven autonomy ensures that spacecraft can perform navigation, control, health monitoring, and fault diagnosis without ground-based oversight. For example, on-board AI systems enable real-time trajectory adjustments, anomaly detection in critical subsystems, and self-healing mechanisms to mitigate failures. These capabilities are vital for long-duration missions (e.g., deep-space exploration) where communication delays and limited bandwidth render Earth-based control infeasible.
  \item \textbf{Remote Sensing and Data Preprocessing:} Spacecraft equipped with AI can process raw sensor data (e.g., hyperspectral imagery (HSI), Synthetic Aperture Radar (SAR) data) \textit{on-board} to prioritize scientifically relevant information. This preprocessing reduces the transmission of redundant or low-value data to Earth, addressing bandwidth constraints. Techniques like feature extraction, noise reduction, and compression are often implemented at the edge, leveraging AI to optimize downstream analysis.
  \item \textbf{Communication Optimization:} AI enables adaptive communication strategies. For instance, AI models can predict optimal communication windows or compress data streams to minimize latency and power consumption, which is critical for resource-constrained missions.
\end{enumerate}   
While AI encompasses diverse techniques (e.g., rule-based systems, evolutionary algorithms), this survey emphasizes NNs due to their scalability, ability to handle high-dimensional data (e.g., multispectral imagery), and suitability for parallelized FPGA implementations. NNs are uniquely positioned to address challenges such as real-time decision-making, adaptive learning in uncertain environments, and efficient resource utilization, all of which are pivotal for next-generation space systems. However, the inherently probabilistic nature of these algorithms requires ensuring their reliability and robustness in safety-critical space applications~\cite{nn_safety_critical,ANN_reliability_Space}.

\subsection{Neural Networks}
\label{sec:background_nn}
Traditional artificial neural networks (ANNs) are grounded in the concept of a perceptron~\cite{mccullochLogicalCalculusIdeas1943}, an abstract model of a biological neuron. This foundational concept paved the way for multi-layer perceptron (MLP)~\cite{popescuMultilayerPerceptronNeural2009} networks, which comprise fully connected (FC) layers. Subsequently, Convolutional Neural Networks (CNNs) emerged~\cite{fukushimaNeocognitronSelforganizingNeural1980}, incorporating convolutional layers. These layers extract features from data, making them highly effective and widely employed in computer vision~\cite{nn_types}. Each connection between neurons is associated with a weight, and each neuron may include a bias; together, they constitute the network's parameters. Commonly represented using 32-bit floating-point (FP32) arithmetic, these parameters can be compressed via quantization~\cite{nn_quantization,gholamiSurveyQuantizationMethods2022} to minimize memory footprint. For instance, quantization can reduce parameters to a single bit, as demonstrated in Binarized Neural Networks (BNNs)~\cite{courbariauxBinarizedNeuralNetworks2016}.

A network's total \textbf{number of operations} is the aggregate across all its layers, where each layer requires specific computations. Equation~\ref{eq:operations_fully_connected} defines the operation count for an FC layer:
\begin{equation}
\label{eq:operations_fully_connected}
\text{Operations} = 2 \times (\text{Input Size}) \times (\text{Output Size})
\end{equation}
Here, the input size represents the number of neurons in the preceding layer, and the output size corresponds to the current layer's neuron count. 

As NNs matured, diverse \textbf{architectures} emerged~\cite{nn_types,wangConvergenceEdgeComputing2020}, including Self-Organizing Maps (SOMs) \cite{SOM}, Graph Neural Networks (GNNs) \cite{GNN}, and Spiking Neural Networks (SNNs)~\cite{maassNetworksSpikingNeurons1997}. SNNs closely mimic the behavior of biological brains. Unlike traditional ANNs, which utilize continuous real-valued activations, SNN neurons communicate via discrete spike events over time. These spiking neurons typically act as leaky integrators, accumulating incoming spikes until reaching a threshold that triggers an emission. Although SNNs are noted for their high power efficiency, they can exhibit elevated latency compared to conventional ANNs~\cite{snn_survey}.

NNs can be tailored for a broad spectrum of tasks by connecting neurons and layers in various configurations. This survey highlights several key \textbf{task types}: classification (predicting a single class label or a probability distribution over multiple classes), segmentation (assigning pixel-wise class labels in an image), object detection (identifying and localizing objects via bounding boxes, class labels, and confidence scores), generative models (producing new content such as images or text), keypoint detection (locating specific coordinate points within an image), depth estimation (generating per-pixel depth values to create a depth map).

Finally, NN \textbf{training} strategies are critical to mission capability. We categorize training methods as on-board or off-board and supervised or unsupervised. On-board training involves updating the NN's weights on the same device that performs inference. In contrast, off-board training occurs when the NN is trained on a separate device before deploying for inference. Supervised learning entails providing expected outputs/labels during training, while unsupervised learning involves training with unlabeled data~\cite{dnn_basics}.

\subsection{FPGA-based Neural Network Accelerators}
\label{sec:FPGA_NN_accelerators}
\begin{figure}[!ht]
  \centering
  \hfill
  \begin{subfigure}[c]{0.45\textwidth}
    \centering
    \includegraphics[width=\textwidth]{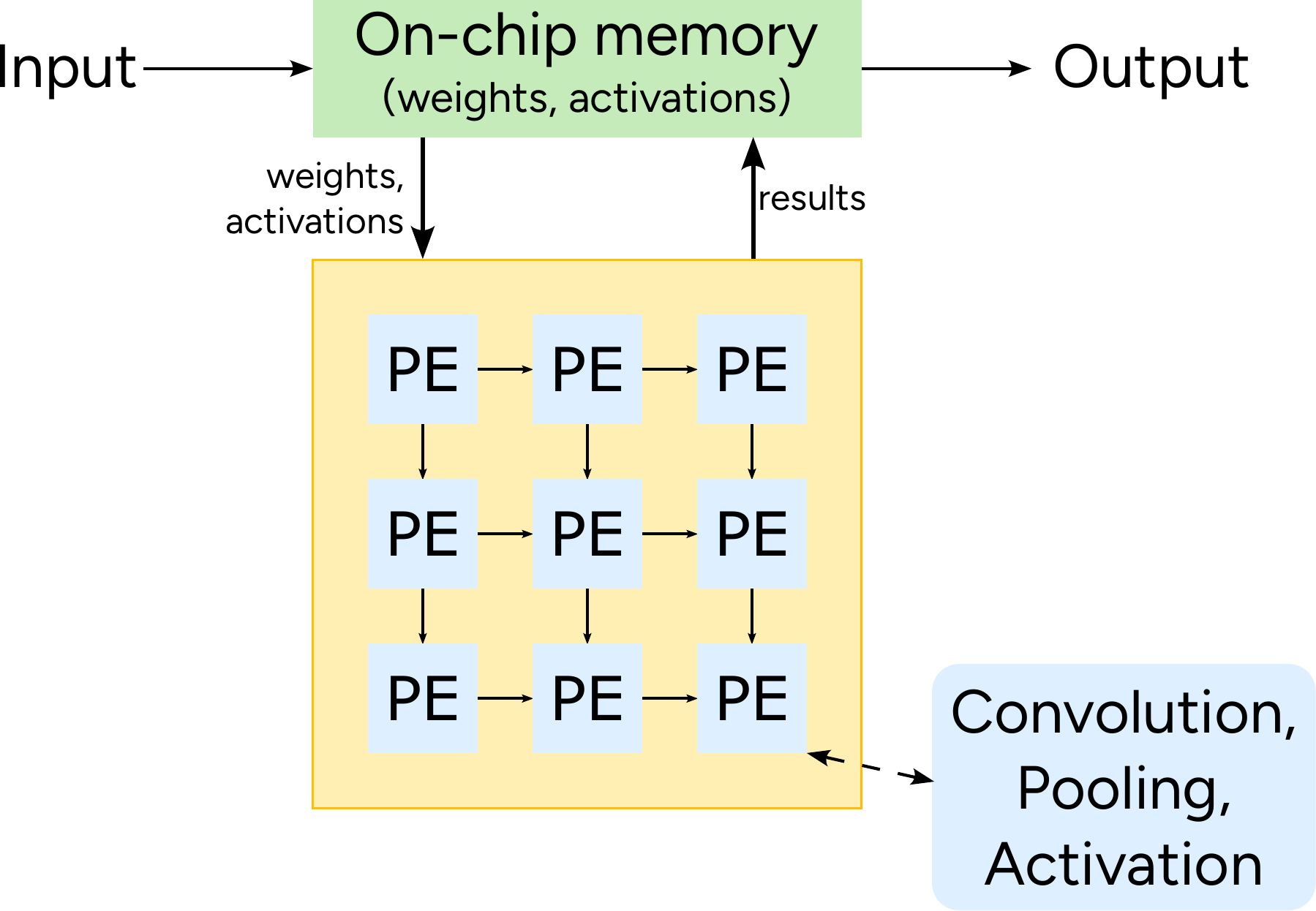}
    \caption{Time-multiplexed PE array with a shared on-chip buffer}
    \label{fig:NN_MPE}
  \end{subfigure}
  \hfill
  \begin{subfigure}[c]{0.45\textwidth}
    \centering
    \includegraphics[width=\textwidth]{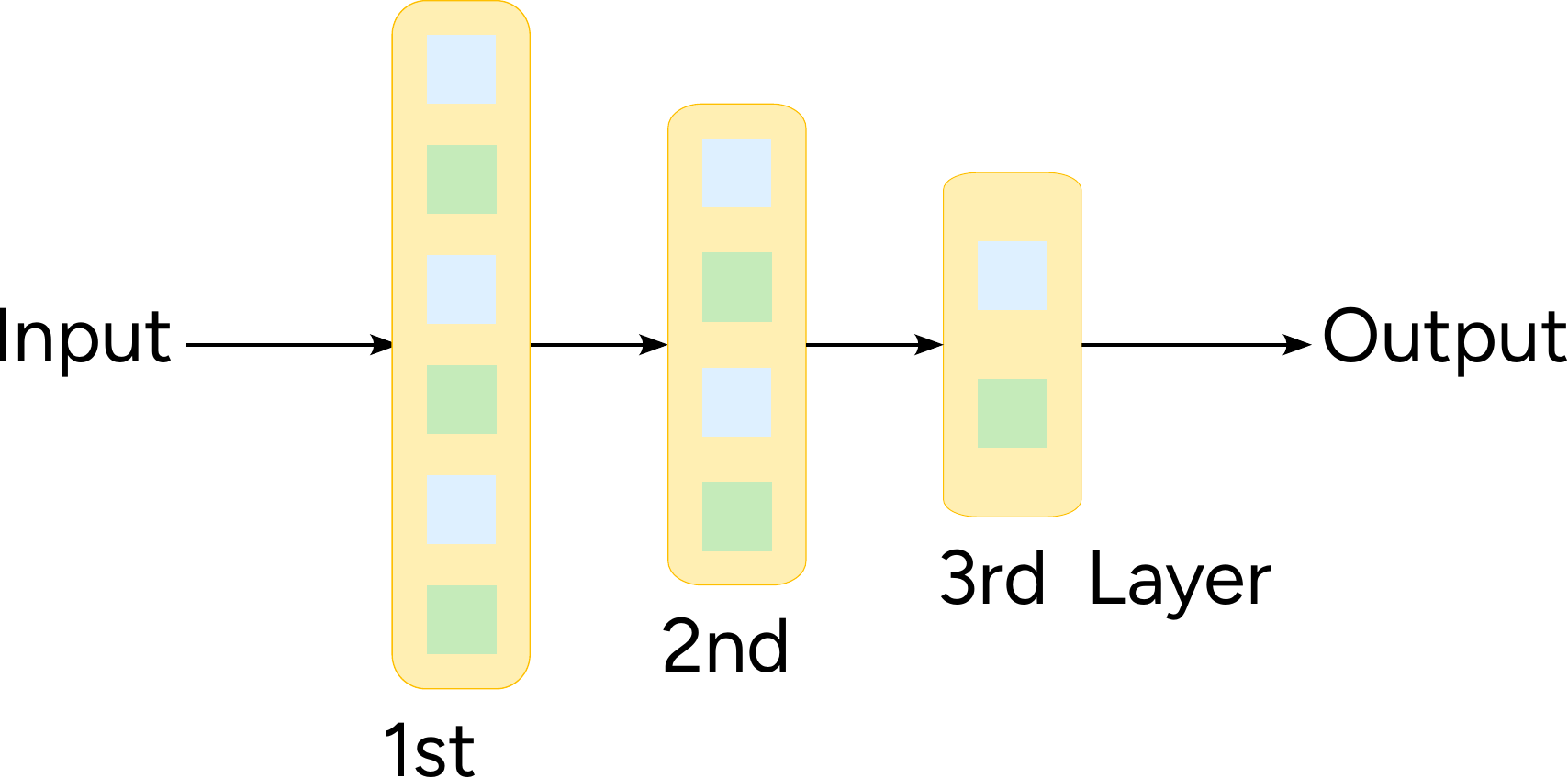}
    \caption{Dataflow example with 3 layers}
    \label{fig:NN_Pipeline_DataFlow}
  \end{subfigure}
  \hfill
  \caption{Architecture of an FPGA-based NN accelerator}
  \label{fig:fpga_nn_accelerator}
  \Description[Architecture of an FPGA-based NN accelerator]{Two FPGA-based NN accelerator paradigms. (a) Time-multiplexed: a PE array is reused across layers; a shared on-chip buffer stores weights and activations, feeds the array, and receives intermediate results for the next layer. Activations flow left to right and weights top to bottom. (b) Dataflow: each layer is mapped to dedicated PEs and memory, forming a pipeline.}
\end{figure}

In this survey, we categorize FPGA-based NN accelerators into two main architectural paradigms: dataflow (Figure~\ref{fig:NN_Pipeline_DataFlow}) and time-multiplexed (Figure~\ref{fig:NN_MPE}). Dataflow architectures spatially map all NN layers onto the FPGA fabric, assigning dedicated processing elements (PEs) and memory blocks to each layer. This design forms a deep pipeline in which data streams sequentially through layer-specific PEs. FINN is an example of this approach~\cite{umurogluFINNFrameworkFast2017}. By contrast, time-multiplexed architectures reuse a set of configurable PEs across multiple NN layers. These designs commonly employ systolic arrays to accelerate matrix multiplication, a fundamental NN operation~\cite{systolic_array_survey}. They compute layers sequentially through time, storing weights and activations in a shared on-chip buffer, as exemplified by the AMD Deep Learning Processor Unit (DPU)~\cite{amdDPUCZDX8GZynqUltraScale2025}. While dataflow accelerators may leverage resources more efficiently, time-multiplexed accelerators may support larger NN models.

Each paradigm may also incorporate elements of the other within its internal components. For instance, a dataflow accelerator might use time-multiplexing at individual layers to conserve resources, whereas a time-multiplexed accelerator could execute groups of layers in a pipeline. Beyond these architectural choices, accelerators can be designed to support multiple NN models, a specific model architecture with reconfigurable weights, or a highly optimized single model with fixed weights.

An accelerator's performance is typically evaluated using inference latency, throughput, and operations per second (OP/s)\footnote{OP/s is sometimes written as OPS in the literature.}. Inference latency refers to the time required for an input to propagate through the entire NN and yield an output. Throughput quantifies the processing rate of these inputs, standardly measured in frames per second (FPS) or samples per second. OP/s is then derived by multiplying this throughput by the total operations executed per inference.

In the context of space missions, power consumption and energy efficiency are equally critical. Power consumption is either measured directly during NN inference or estimated using RTL analysis tools~\cite{rtl_power_estimation}. Energy efficiency, defined as the energy consumed per inference, bridges hardware architectural choices, which dictate power, with algorithmic efficiency, which dictates latency. Furthermore, operations per second per watt (OP/s/W) is a vital metric for general-purpose hardware. By abstracting performance from specific use cases, it effectively highlights the suitability of FPGA-based NN accelerators for resource-constrained deployments.

\subsection{Fault Tolerance and Robustness}
\label{sec:fault_tolerance}
\begin{wrapfigure}{r}{0.5\textwidth}
\centering
\includegraphics[width=\linewidth]{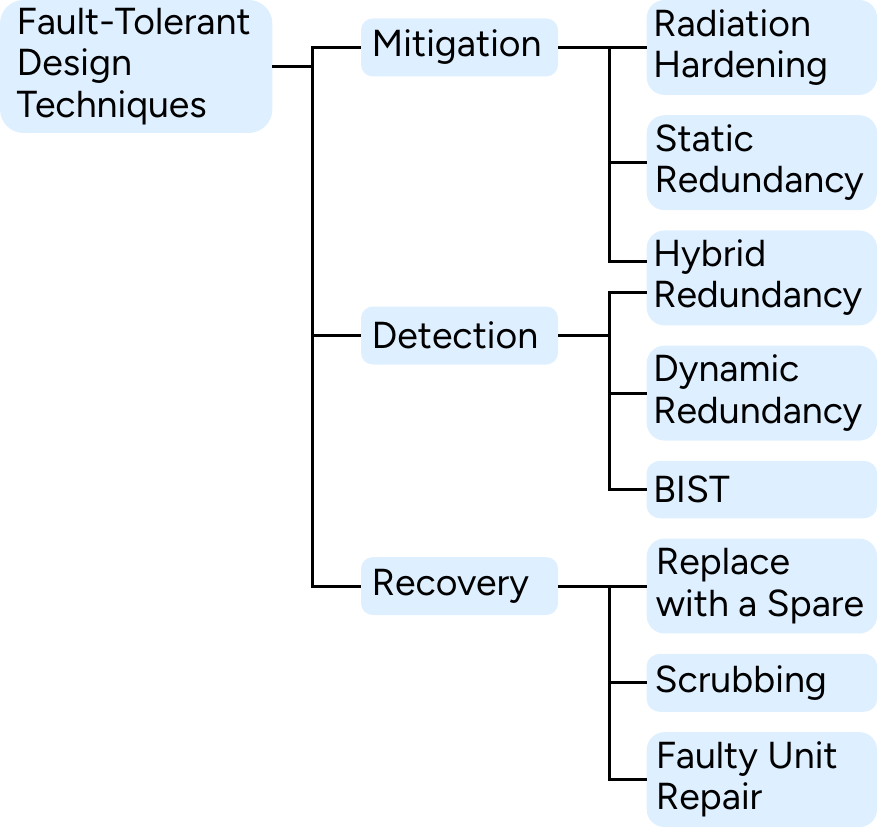}
\caption{\label{fig:Fault_Tolerant_Design_summary}Different fault tolerance design techniques}
\Description[Taxonomy of fault-tolerance techniques]{A diagram grouping fault-tolerance techniques into fault mitigation, fault detection, and fault recovery. Mitigation includes radiation-hardened devices and static redundancy such as TMR. Detection includes dynamic redundancy, duplication with comparison, and BIST. Recovery includes repair, replacement, and FPGA configuration-memory scrubbing.}
\end{wrapfigure}
Fault-tolerance strategies fall into three main categories (Figure~\ref{fig:Fault_Tolerant_Design_summary}): (i) fault mitigation, (ii) fault detection, and (iii) fault recovery~\cite{yarzadaBriefSurveyFault2022}. \textbf{Fault mitigation} seeks to prevent radiation effects from propagating. Examples include using radiation-hardened or radiation-tolerant devices that withstand specific TID levels~\cite{barnabyTotalIonizingDoseEffectsModern2006,kernsDesignRadiationhardenedICs1988} and static (passive) redundancy schemes such as TMR~\cite{Fault_Tolerance_Elena}. TMR masks faults via triplication and majority voting. \textbf{Fault detection} identifies SEEs and can be achieved using dynamic techniques (active redundancy)~\cite{Fault_Tolerance_Elena}. One such technique is duplication with comparison, which raises an error alarm when the outputs mismatch. Another fault-detection method is built-in self-test (BIST)~\cite{abramoviciBISTbasedTestDiagnosis2001}, which verifies outputs against known responses (typically pausing full operation). Roving error detection~\cite{abramoviciRovingSTARsIntegrated2001} mitigates this pause by running BIST on individual segments, allowing the rest of the system to operate and enabling precise localization. Hybrid redundancy combines mitigation and detection (e.g., N-modular redundancy with spares or self-purging with N-redundant modules)~\cite{Fault_Tolerance_Elena}. \textbf{Fault recovery} restores correct system behavior by repairing or replacing faulty units and, for FPGAs, by applying configuration memory scrubbing~\cite{nidhinReviewSEUMitigation2018}. Scrubbing periodically reads, verifies, and rewrites the configuration bitstream; blind, read-back, and hybrid scrubbing differ in timing and trigger conditions, and they may target the full configuration memory or only affected regions~\cite{heinerFPGAPartialReconfiguration2009}.

These techniques cannot be applied blindly to SRAM-based FPGAs. Unlike a bitflip in a CPU or ASIC register, an SEU in the FPGA configuration memory---which encodes LUT functions, programmable routing, and I/O settings---persists until the bitstream is rewritten and can silently change the circuit's logic or routing~\cite{fpga_reliability,nidhinReviewSEUMitigation2018}. Because configuration memory constitutes the vast majority of device state, it is the dominant failure mode of SRAM FPGAs in radiation environments, whereas antifuse and flash FPGAs, whose configuration cells are largely SEU-immune, present a reliability challenge similar to ASICs~\cite{fpga_reliability}. Achieving high reliability on SRAM FPGAs therefore requires pairing frequent configuration scrubbing with spatial redundancy such as TMR~\cite{fpga_reliability,heinerFPGAPartialReconfiguration2009}.

Mitigation, detection, and recovery exploit redundancy across hardware, data, and software, as well as across space and time. For a comprehensive view of traditional fault-tolerant methods, we refer the reader to Elena Dubrova's book \textit{Fault-Tolerant Design}~\cite{Fault_Tolerance_Elena}. Beyond these methods, NN-specific approaches are emerging. NNs can be trained in a fault-aware manner by injecting faults into intermediate feature maps during training so that their parameters adapt to tolerate such faults~\cite{nn_fault_aware}. At the algorithmic level, networks can use ``smart'' layers (for example, pooling layers that validate value ranges before propagation) or compare sequential outputs at inference time to detect unexpected deviations. An accelerator's reliability can be evaluated through fault-injection and beam experiments; Paolo Rech provides a survey on this topic~\cite{ANN_reliability_Space}. Gao et al.~\cite{cnn_see_evaluation} evaluated the AMD DPU through radiation beam experiments on Zynq-7000 SoCs. They noticed that SEEs that caused system corruption predominantly affected the data mover and instruction scheduler, whereas SEEs that degraded accuracy primarily affected the computing engine.

Safety-critical applications require not only reliability but also robustness~\cite{nn_safety_critical}. Some fault-tolerance concepts may be adapted to meet robustness requirements, ensuring correct actions under faults or uncertainty. Gleirscher et al. \cite{nn_robustness} propose a quantitative hazard-rate approach and an architecture composed of three 2-out-of-2 modules, each implemented with two statistically independent object-detection systems operating together with a voter in a 3-out-of-3 configuration. This architecture, applied to camera-based object detection, meets standards relevant to autonomous railway systems (e.g., ANSI/UL 4600, ISO 21448, and EN 5012x).

\section{Literature Review}
\label{sec:Section_3}
This section summarizes the reviewed literature, focusing on studies published between 2014 and 2024 that target space-use cases and have a concrete FPGA implementation. We applied the same search string (see Listing~\ref{lst:string_web_of_science}) across Web of Science, Scopus, the IEEE Digital Library, and the ACM Digital Library to identify relevant works. We then cross-referenced these results with citations from the gathered papers and supplementary Google Scholar searches, systematically filtering the final corpus to align with our survey's objectives.

\begin{lstlisting}[caption=String used to search in the Web of Science, label=lst:string_web_of_science, breaklines]
(("fpga*" OR "field-programmable gate array*" OR "field programmable gate array*" OR "programmable logic") AND ("ai" OR "mlp" OR "cnn" OR "perceptron" OR "deep learning" OR "deep-learning" OR "spiking network**" OR "neuromorphic" OR "Neur* Network**" OR "machine learning" OR "ML" OR "artificial intelligence") AND ("spacecraft*" OR "space mission*" OR "satellite*" OR "space app*" OR "cubesat*"))
\end{lstlisting}

We structured each summary into two paragraphs. The first outlines the core research problem and proposed solution, detailing methodological aspects such as the NN architecture, dataset, quantization strategy, accelerator design, and FPGA implementation. The second paragraph evaluates the reported results, discussing task accuracy alongside key hardware metrics.

To systematically evaluate this literature, we extracted key metrics and classified them into two primary domains: algorithm-related and hardware-related. The former captures the intrinsic properties of the neural network models, datasets, and training procedures, providing insight into computational demands and task accuracy. The latter focuses on the physical FPGA implementation, assessing the resource footprint, efficiency, power profile, and radiation tolerance of the accelerators, constraints that are critically restrictive in space missions. Appendix~\ref{tab:evaluation_metrics} details these precise metrics and the rationale for their inclusion.

We recommend two approaches for navigating this survey: readers may proceed sequentially through the detailed summaries, or they may bypass them and jump directly to Section~\ref{sec:Section_4} for a discussion of the literature. For quick reference, the quantitative data extracted from all reviewed works are consolidated in the appendices at the end of this paper (Appendix~\ref{app:algorithm_table} for algorithm-related metrics and Appendix~\ref{app:hardware_table} for hardware-related metrics). Finally, we provide a GitHub repository containing the complete dataset and the scripts used to generate data plots~\footnote{\url{https://github.com/asap-space/SPACE_FPGA_NNs_SURVEY_DATA}}.

We organized the remainder of this section into three subsections based on the target application domain. We begin with Earth Observation (EO) and remote sensing applications, which represent the majority of the reviewed works. We then cover spacecraft perception, autonomy, and navigation, and conclude with anomaly detection and undefined use cases.

\subsection{Earth Observation and Remote Sensing Applications}
Given the spatial nature of the data in this section, the majority of the proposed accelerators employ convolutional layers for feature extraction. Notable exceptions utilizing alternative architectures include: Zhuang and Low (2014)~\cite{zhuangRealTimeRunway2014}, who implemented a pulse-coupled neural network (PCNN); Jain et al. (2018)~\cite{jainExplorationFPGABased2018}, who utilized an MLP; Lemaire et al. (2020)~\cite{lemaireFPGAbasedHybridNeural2020} and Abderrahmane et al. (2022)~\cite{abderrahmaneSPLEATSPikingLowpower2022}, who explored SNNs; and Zhang et al. (2022)~\cite{zhangAccurateLowlatencyEfficient2022}, who implemented a GNN.

Regarding training methodologies, most of these works rely on off-board supervised learning. Deviations from this trend include Mazouz and Nguyen (2024)~\cite{mazouzOnlineContinualStreaming2024}, who proposed an on-board continuous learning approach, and Castelino et al. (2024)~\cite{castelinoEnergyEfficientArtefactDetection2024}, who employed unsupervised off-board training. Furthermore, accelerators capable of supporting multiple NN architectures include the AMD DPU, as well as custom architectures developed by Yan et al. (2022)~\cite{yanAutomaticDeploymentConvolutional2022}, Ni et al. (2023)~\cite{niAlgorithmHardwareCoOptimizationDeployment2023}, and Shao et al. (2024)~\cite{shaoConfigurableAcceleratorCNNBased2024}. All studies that leverage the AMD DPU quantize their NN parameters to 8-bit integers using Vitis AI. Finally, among the reviewed literature, Sabogal et al. (2019~\cite{sabogalReCoNReconfigurableCNN2019}, 2021~\cite{sabogalReconfigurableFrameworkResilient2021}) uniquely address hardware reliability by integrating fault-tolerant mechanisms and evaluating their accelerator through beam testing. Table~\ref{tab:eo_and_rs} summarizes additional key implementation distinctions across these studies.

\begin{table*}[htbp]
  \centering
  \small
  \caption{Distinctive characteristics of each earth observation and remote sensing system reviewed in this section}
  \label{tab:eo_and_rs}
  \footnotesize
  \renewcommand{\arraystretch}{1.2}
  \begin{tabular}{llll}
    \toprule
    \textbf{Key.} & \makecell[tl]{\textbf{Task}\\\textbf{Type}} & \makecell[tl]{\textbf{FPGA}\\\textbf{Family}}   & \makecell[tl]{\textbf{Hardware}\\\textbf{Design}} \\
    \midrule
    \textbf{\hyperref[par:zhuang2014]{Zhuang and Low (2014)~\cite{zhuangRealTimeRunway2014}}} & Segmentation & Spartan 3A & -- \\
    \textbf{\hyperref[par:chen2018]{Chen et al. (2018)~\cite{chenFPGABasedImplementation2018}}} & Classification & -- & HDL \\
    \textbf{\hyperref[par:jain2018]{Jain et al. (2018)~\cite{jainExplorationFPGABased2018}}} & Classification & Spartan 6 & -- \\
    \textbf{\hyperref[par:hashimoto2018]{Hashimoto et al. (2018)~\cite{hashimotoShipClassificationSAR2018}}} & Classification & Zynq 7000 SoC & FINN \\
    \textbf{\hyperref[par:pitsis2019]{Pitsis et al. (2019)~\cite{pitsisEfficientConvolutionalNeural2019}}} & Classification & Zynq UltraScale+ MPSoC & -- \\
    \textbf{\hyperref[par:sabogal2019]{Sabogal et al. (2019~\cite{sabogalReCoNReconfigurableCNN2019}, 2021~\cite{sabogalReconfigurableFrameworkResilient2021})}} & Segmentation & \makecell[tl]{Zynq 7000 SoC \& \\ Zynq UltraScale+ MPSoC} & HLS \\
    \textbf{\hyperref[par:liu2019]{Liu and Luk (2019)~\cite{liuEfficientAcceleratorDNNBased2019}}} & Segmentation & Arria 10 SoC & Verilog HDL \\
    \textbf{\hyperref[par:li2019]{Li et al. (2019)~\cite{liEfficientObjectDetection2019}}} & Object Detection & Zynq 7000 SoC & Verilog HDL \\
    \textbf{\hyperref[par:bahl2019]{Bahl et al. (2019)~\cite{bahlLowpowerNeuralNetworks2019}}} & Segmentation & Cyclone V & VGT \\
    \textbf{\hyperref[par:reiter2020]{Reiter et al. (2020)~\cite{reiterFPGAAccelerationQuantized2020}}} & Classification & Zynq 7000 SoC & FINN \\
    \textbf{\hyperref[par:lemaire2020]{Lemaire et al. (2020)~\cite{lemaireFPGAbasedHybridNeural2020}}} & Classification & Cyclone V & VHDL \\
    \textbf{\hyperref[par:zhang2021]{\makecell[tl]{Zhang et al. (2021)~\cite{zhangFPGAImplementationCNNBased2021},\\ Yan et al. (2022)~\cite{yanAutomaticDeploymentConvolutional2022}, and\\ Ni et al. (2023)~\cite{niAlgorithmHardwareCoOptimizationDeployment2023}}}} & \makecell[tl]{Object Detection \&\\ Classification} & \makecell[tl]{Zynq 7000 SoC,\\ Artix 7 \& Virtex 7} & \makecell[tl]{VHDL \&\\ SystemVerilog} \\
    \textbf{\hyperref[par:rapuano2021]{\makecell[tl]{Rapuano et al. (2021)~\cite{rapuanoFPGABasedHardwareAccelerator2021} and\\ Pacini et al. (2021)~\cite{paciniMultiCacheSystemOnChip2021}}}} & Classification & Zynq UltraScale+ MPSoC & VHDL \\
    \textbf{\hyperref[par:yang2022]{Yang et al. (2022)~\cite{yangAlgorithmHardwareCodesign2022}}} & Object Detection & Virtex 7 & HLS \\
    \textbf{\hyperref[par:kyriakos2022]{Kyriakos et al. (2022)~\cite{kyriakosResourcesPowerEfficient2022}}} & Classification & Virtex 7 & VHDL \\
    \textbf{\hyperref[par:neris2022]{Neris et al. (2022)~\cite{nerisFPGABasedImplementationCNN2022}}} & Object Detection & Kintex UltraScale & HLS \\
    \textbf{\hyperref[par:pitonak2022]{Pitonak et al. (2022)~\cite{pitonakCloudSatNet1FPGABasedHardwareAccelerated2022}}} & Classification & Zynq 7000 SoC & FINN \\
    \textbf{\hyperref[par:zhang2022]{Zhang et al. (2022)~\cite{zhangAccurateLowlatencyEfficient2022}}} & Classification & Zynq UltraScale+ MPSoC & -- \\
    \textbf{\hyperref[par:abderrahmane2022]{Abderrahmane et al. (2022)~\cite{abderrahmaneSPLEATSPikingLowpower2022}}} & Segmentation & Cyclone V & VHDL \\
    \textbf{\hyperref[par:wang2022]{Wang et al. (2022)~\cite{wangHardwareAccelerationImplementation2022}}} & Object Detection & Virtex 7 & Verilog HDL \\
    \textbf{\hyperref[par:papatheofanous2022]{Papatheofanous et al. (2022)~\cite{papatheofanousSoCFPGAAcceleration2022}}} & Segmentation & Zynq UltraScale+ MPSoC & AMD DPU \\
    \textbf{\hyperref[par:zhao2023]{Zhao et al. (2023)~\cite{zhaoHardwareAccelerationSatellite2023}}} & Object Detection & Zynq UltraScale+ MPSoC & AMD DPU \\
    \textbf{\hyperref[par:mazouz2024]{Mazouz and Nguyen (2024)~\cite{mazouzOnlineContinualStreaming2024}}} & Object Detection & Zynq 7000 SoC & -- \\
    \textbf{\hyperref[par:shao2024]{Shao et al. (2024)~\cite{shaoConfigurableAcceleratorCNNBased2024}}} & Object Detection & Virtex 7 & Verilog HDL \\
    \textbf{\hyperref[par:castelino2024]{Castelino et al. (2024)~\cite{castelinoEnergyEfficientArtefactDetection2024}}} & Generative & Zynq UltraScale+ MPSoC & AMD DPU \\
    \textbf{\hyperref[par:zhang2024]{Zhang et al. (2024)~\cite{zhangEnergyefficientDehazingNeural2024}}} & Generative & Zynq 7000 SoC & Verilog HDL \\
    \textbf{\hyperref[par:cratere2024]{Cratere et al. (2024)~\cite{cratereEfficientFPGAacceleratedConvolutional2024}}} & Segmentation & Zynq 7000 SoC & AMD DPU \\
    \textbf{\hyperref[par:kimOn2024]{Kim et al. (2024)~\cite{kimOnOrbitAICloud2024}}} & \makecell[tl]{Classification \&\\ Segmentation} & Zynq 7000 SoC & HLS \\
    \bottomrule
  \end{tabular}
\end{table*}

\phantomsection\label{par:zhuang2014}
\textbf{Zhuang and Low (2014)~\cite{zhuangRealTimeRunway2014}} proposed a PCNN accelerator for airfield runway detection in 400x400 RGB images on-board satellites. The PCNN performed image segmentation, and its output was filtered using the Hough transform to detect runway features. They implemented it on a Spartan 3A FPGA.

They compared PCNN's performance with a conventional seeded region growing method~\cite{shihAutomaticSeededRegion2005}. This comparison revealed that the proposed method outperformed the conventional method by two or three orders of magnitude, depending on the number of segments processed. Furthermore, when detecting runways, this implementation achieved true-positive and true-negative rates of 88\% and 90\%, respectively.

\phantomsection\label{par:chen2018}
\textbf{Chen et al. (2018)~\cite{chenFPGABasedImplementation2018}} proposed an FPGA-based CNN accelerator for classifying HSI pixels into predefined ground-material categories. This accelerator dataflow design incorporated convolution, batch normalization, and dense layers, and it stored layer parameters on chip to reduce memory-access latency. The authors trained and tested it on the Pavia University dataset.

This system achieved 94.58\% overall accuracy and 92.24\% average accuracy across nine classes. These results were comparable to GPU execution. Its processing speed scaled with FPGA logic utilization, reaching 999 pixels per second, or 208 seconds per image.

\phantomsection\label{par:jain2018}
\textbf{Jain et al. (2018)~\cite{jainExplorationFPGABased2018}} proposed an MLP accelerator for on-board payload data processing that leveraged the Residue Number System for number representation. They claimed it required fewer hardware resources and yielded results equivalent to those of fixed- or floating-point arithmetic. This system was tested on the Spartan-6 FPGA using the Crowdsourced Mapping Dataset \cite{johnsonCrowdsourcedMapping2016, johnsonIntegratingOpenStreetMapCrowdsourced2016} at a clock frequency of 100 MHz.

This system consumed 3.1 W, had an accuracy of 82.46\%, and a latency of 7.81 microseconds ($\mu s$). Notably, compared to an Intel i7 6700HQ CPU, which achieved 86.80\% accuracy and 3.67 $\mu s$ latency when running the same NN, this system exhibited higher energy efficiency without significantly compromising performance.

\phantomsection\label{par:hashimoto2018}
\textbf{Hashimoto et al. (2018)~\cite{hashimotoShipClassificationSAR2018}} proposed a ship detection and classification method utilizing CNN models for maritime/sea-only scenes. These models were trained and evaluated on SAR images acquired by the Phased Array type L-band SAR on board the Advanced Land Observing Satellite. Their FPGA implementation leveraged FINN \cite{umurogluFINNFrameworkFast2017} to generate a CNN accelerator, and was tested using 36 SAR images containing 513 ships.

The GPU-based implementation achieved 99.8\% accuracy in ship detection and 65.7\% in type classification, with an average error margin of 26\% for ship length classification. In contrast, the FPGA-based implementation successfully detected all 513 ships, achieving 100\% accuracy on this test dataset. Notably, the processing time for a single image in the FPGA-based implementation was 330 milliseconds (ms), outperforming the GPU and CPU implementations, which took 1165 and 7381 ms, respectively.

\phantomsection\label{par:pitsis2019}
\textbf{Pitsis et al. (2019)~\cite{pitsisEfficientConvolutionalNeural2019}} proposed a CNN-based accelerator to estimate galaxy redshift, which is a prerequisite for calculating the acceleration of the universe. They implemented this accelerator on the Quad-FPGA Daughter-Board (QFDB) and evaluated it using simulated spectroscopic measurements from the ESA Euclid deep-space mission~\cite{raccaEuclidMissionDesign2016}. To deploy the CNN algorithm~\cite{stivaktakisConvolutionalNeuralNetworks2020} on board, they reduced its memory footprint through weight pruning, quantization, and weight clustering. These techniques reduced the model size from 173 MB to 11 MB, corresponding to a 16x compression factor.

This system computed 4334 signals per second while consuming a maximum of 109 Joules. Compared with an Nvidia Quadro P1000 GPU, two QFDBs operating in parallel achieved 1.23x higher throughput, consumed 5x less energy, and exhibited one order of magnitude lower latency. Furthermore, the system achieved 416 GFLOP/s and 99\% classification accuracy.

\phantomsection\label{par:sabogal2019}
\textbf{Sabogal et al. (2019~\cite{sabogalReCoNReconfigurableCNN2019}, 2021~\cite{sabogalReconfigurableFrameworkResilient2021})} proposed and subsequently enhanced the Reconfigurable ConvNet (RECON), a CNN accelerator designed for on-board semantic segmentation using a SegNet-based model~\cite{badrinarayananSegNetDeepConvolutional2015}. To ensure fault tolerance, RECON leveraged the Hybrid Adaptive Reconfigurable Fault Tolerance (HARFT)~\cite{wilsonHybridAdaptiveReconfigurable2017} framework, which mitigated SEEs through configuration memory scrubbing and TMR. In their 2021 enhancement, the authors implemented dynamic runtime configuration for NN weights and TMR utilization, and replaced the line-buffer approach with a more general systolic-array approach. Moreover, they included optimizations such as int8 model quantization, feature map caching, Winograd transformations, and layer fusion. The authors trained this system on the Potsdam dataset~\cite{potsdam20182d} and deployed it on a Zynq SoC and a Zynq UltraScale+ MPSoC. They tested its reliability through fault injection and neutron irradiation.

There were clear trade-offs between performance, power consumption, and reliability across different configurations. Notably, in the authors' first work, the fastest inference time was 0.7 s on the UltraScale+ MPSoC using the Q9.18 fixed-point representation, yielding an accuracy of 88.01\% and a power consumption of 25.3 W (21.6 W idle power + 3.7 W dynamic power). Additionally, the lowest power consumption was 8.73 W (7.13 W idle power + 1.60 W dynamic power) on the Zynq SoC using the Q9.16 fixed-point representation, resulting in 88.00\% accuracy and an inference time of 4.6 s. In their second work, they improved performance while degrading mean intersection-over-union (IoU) by 1.68\% or less and F1 accuracy by 1.33\% or less. Notably, when comparing configurations with and without TMR, the TMR-enabled configuration exhibited a 56-fold reduction in vulnerability to SEE-induced application hangs, despite increased FPGA resource utilization, lower performance, and reduced power efficiency.

\phantomsection\label{par:liu2019}
\textbf{Liu and Luk (2019)~\cite{liuEfficientAcceleratorDNNBased2019}} proposed a CNN accelerator for remote sensing image segmentation, integrating convolutional and deconvolutional layers within a single module. Their accelerator leveraged parallelism across channels and filters to enhance computational efficiency, and incorporated several optimizations into its design, including 8-bit quantization, layer fusion, input reshaping, and tailored DSP configuration. The authors evaluated it using 256x256-pixel urban surface images on an Arria 10 SX SoC, using a modified U-Net model as the test case. 

The proposed U-Net achieved 80.1\% accuracy, with an inference time of 17.4 ms and a throughput of 1578 GOP/s. When running on the FPGA, this system consumed 32 W.

\phantomsection\label{par:li2019}
\textbf{Li et al. (2019)~\cite{liEfficientObjectDetection2019}} proposed a Context-Based Feature Fusion Single-Shot multi-box Detector and implemented it on a Zynq 7000 FPGA for object detection in high-resolution satellite images. Building on the SSD architecture~\cite{liuSSDSingleShot2016}, this model uses MobileNet~\cite{howardMobileNetsEfficientConvolutional2017} as its backbone to improve efficiency. The Deep Learning Processor (DLP) included a Neural Processing Engine (NPE), control modules, and memory buffers; this NPE is an array of Neural Processing Units (NPUs), each containing a matrix of PEs. They fine-tuned a model pre-trained on VOC0712~\cite{everinghamPascalVisualObject2010} using the NWPU VHR-10 dataset~\cite{chengSurveyObjectDetection2016, chengLearningRotationInvariantConvolutional2016} to evaluate performance.

This algorithm achieved 91.42\% mean Average Precision (mAP) while the DLP consumed 19.52 W, making it 29.74x and 2.87x more efficient than an Intel i7-7700 CPU and an NVIDIA GeForce GTX1070T GPU, respectively. Although it did not exceed other FPGA-based accelerators in raw performance, this architecture achieved the highest performance density (1.97 OP/DSP/cycle).

\phantomsection\label{par:bahl2019}
\textbf{Bahl et al. (2019)~\cite{bahlLowpowerNeuralNetworks2019}} proposed several compact NN architectures, namely C-UNet, C-UNet++, C-FCN, and C-FCN++, for semantic segmentation on low-power devices. These networks were trained and evaluated for two use cases: cloud and forest segmentation. The datasets utilized for these use cases were the Cloud-38~\cite{mohajeraniCloudNetEndEndCloud2019} and CloudPeru2 datasets~\cite{moralesCloudDetectionHighResolution2018} for cloud segmentation and the EOLearn Slovenia 2017 dataset~\cite{druschSentinel2ESAOptical2012} for forest segmentation. The C-FCN++ architecture was implemented on a Cyclone V FPGA using VGT~\cite{hamdanVhdlGeneratorHigh2017}. This architecture was the smallest model introduced, with only 273 parameters, occupying 0.047 MB. In contrast, the largest model developed was the C-UNet, with 51,113 parameters, occupying 0.735 MB. 

The performance of these models differed significantly. The C-UNet demonstrated comparable accuracy, precision, and recall to the U-Net model. Conversely, the C-FCN++ exhibited decreases in accuracy of 2.55\%, 5.05\%, and 5.93\% on the Cloud-38, CloudPeru2, and EOLearn Slovenia 2017 datasets, respectively, compared to the C-UNet. The FPGA implementation enabled real-time analysis of images captured by the OPS-SAT's camera, with cloud segmentation performed on a 2048x1944 image in less than 150 ms.

\phantomsection\label{par:reiter2020}
\textbf{Reiter et al. (2020)~\cite{reiterFPGAAccelerationQuantized2020}} proposed a BNN accelerator for cloud detection. Its weights and biases are quantized to 1-bit (binary). This BNN architecture was based on the CNV-W1A1 model in the Brevitas libraries~\cite{pappalardoXilinxBrevitas2023} and implemented on the Artix-7 FPGA using FINN~\cite{umurogluFINNFrameworkFast2017}. The training and testing datasets employed were a custom ESA Copernicus Sentinel-2 dataset~\cite{druschSentinel2ESAOptical2012} and the Planet dataset~\cite{siroshPlanetScaleLandCover2018}, with TMR being discussed as a potential method to mitigate SEE effects. However, its implementation was limited by FPGA memory constraints.

When executed on the FPGA, the authors found the NN model's performance was suboptimal compared to its execution without quantization on both the CPU and the GPU. Specifically, the inference accuracy on the FPGA was 64\%, whereas it reached 91.8\% on the CPU and 91.9\% on the GPU. Despite the relatively low accuracy, the experiment revealed that the BNN model on the FPGA consumed 2.4 W, ran in 2.8 ms, and achieved a throughput of 358.1 images per second. Compared to the Nvidia Tesla M60 GPU, the FPGA demonstrated a significant advantage in speed (7.9 times faster) and energy efficiency (120 times lower power consumption).

\phantomsection\label{par:lemaire2020}
\textbf{Lemaire et al. (2020)~\cite{lemaireFPGAbasedHybridNeural2020}} proposed an accelerator for a Hybrid Neural Network (HNN) designed for cloud detection on-board satellites, leveraging the LeNet model~\cite{lecunGradientbasedLearningApplied1998} as its foundation. This HNN integrated a traditional CNN with an SNN, and its accelerator comprised three primary components. These components were the feature extraction module, implemented following a traditional CNN implementation~\cite{hamdanVhdlGeneratorHigh2017}; the interface between the traditional CNN and SNN component, which converted rate-based data into spike trains; and the classification SNN layers, in which each neuron followed the Integrate and Fire (IF) model~\cite{burkittReviewIntegrateandfireNeuron2006}.

This system was successfully implemented at a clock frequency of 100 MHz on a Cyclone-V FPGA, mirroring the hardware configuration employed by the OPS-SAT~\cite{evansOPSSATESANanosatellite2014,evansOPSSATOperationalConcept2016}. The implementation achieved an accuracy of 87\%, a latency of 43~$\mu\text{s}$, utilized 59\% of logic cells, and consumed 1192.66 mW of power. In contrast, its counterpart, implemented without incorporating SNN components, achieved an accuracy of 88\%, a latency of 25~$\mu\text{s}$, logic cell occupation of 71\%, and power consumption of 1248.44 mW.

\phantomsection\label{par:zhang2021}\label{par:yan2022}\label{par:ni2023}
\textbf{Zhang et al. (2021)~\cite{zhangFPGAImplementationCNNBased2021}, Yan et al. (2022)~\cite{yanAutomaticDeploymentConvolutional2022}, and Ni et al. (2023)~\cite{niAlgorithmHardwareCoOptimizationDeployment2023}} developed related CNN accelerators for optical remote-sensing image processing. Zhang et al. proposed a 32-PE accelerator for an improved YOLOv2, with layer fusion and hybrid quantization: 8-bit fixed-point feature maps and weights, 32-bit fixed-point biases and convolutional outputs, and FP32 batch normalization and activations. They implemented this design on a Zynq 7000 SoC. Yan et al. and Ni et al. independently built on this quantization-and-fusion scheme to support multiple CNN models. Yan et al. proposed a compilation toolchain that automatically converts PyTorch CNN models into hardware instructions for eight reconfigurable PEs on an Artix 7 FPGA, adding unified layer operations, convolution dataflow rearrangement, and dynamic slicing of input feature maps. Ni et al. designed a configurable 32-PE array with an instruction queue on a Virtex-7 FPGA, adding operator fusion and depth-first mapping. All three works evaluated the improved YOLOv2 on the DOTA dataset~\cite{xiaDOTALargescaleDataset2019}. Yan et al. and Ni et al. additionally evaluated scene classification on NWPU-RESISC45~\cite{chengRemoteSensingImage2017}, using an improved VGG16 and VGG16 plus ResNet-34~\cite{simonyanVeryDeepConvolutional2015,heDeepResidualLearning2015}, respectively.

Zhang et al. reported 3.4 s inference at 5.96 W. Yan et al. reported 17.12 s for YOLOv2 and 1.78 s for VGG16 at 3.407 W, and Ni et al. reported 981.4 ms, 89.1 ms, and 40.2 ms for YOLOv2, VGG16, and ResNet-34 at 14.97 W (12.18--25.83 GOP/s/W). Accuracy remained close to the CPU and GPU results, with mAP drops of 0.18\% and 0.2\% for Zhang et al. and Yan et al., and a 0.05\% overall-accuracy drop for Yan et al.'s VGG16. All three FPGAs were more energy efficient than the CPU and GPU baselines: Zhang et al. were 33.4 times more energy efficient than a CPU and used 2.384\% of GPU power, while the GPU remained 47.3 times faster; Yan et al. used 31 times and 73 times less power than an Intel Xeon Gold 5120T and an Nvidia Titan Xp; Ni et al. exceeded an Intel Xeon E5-2697v4 (0.37--1.47 GOP/s/W) and matched or exceeded a Titan Xp (2.45--23.16 GOP/s/W) in energy efficiency.

\phantomsection\label{par:rapuano2021}\label{par:pacini2021}
\textbf{Rapuano et al. (2021)~\cite{rapuanoFPGABasedHardwareAccelerator2021}, and Pacini et al. (2021)~\cite{paciniMultiCacheSystemOnChip2021}} proposed a CNN accelerator for cloud detection on-board satellites, based on a modified CloudScout~\cite{giuffridaCloudScoutDeepNeural2020} model. Rapuano et al. applied layer-wise custom arithmetic precision, reducing the memory footprint by 48\%, from 204 to 107 Mbits, with a 0.3\% drop in accuracy. Their design comprised a custom cache, a shared convolutional PE, an on-chip filter ROM, a max pooling PE, and an FC PE. It was implemented on a Zynq UltraScale+ ZCU106 board and synthesized for a Radiation Tolerant (RT) Kintex UltraScale XQRKU060 FPGA. Pacini et al. then improved the L1 Cache, which reused input feature map data to reduce the input buffer, and the Filter Cache, which stored a single layer's filters on-chip. They demonstrated this design on several FPGAs, including ZU3EG, Z7030, XC7A200T, KU025, and A10 GX 270.

On the Zynq board, Rapuano et al. reported 141.68 ms inference time and 3.4 W, versus an estimated 264.72 ms and 1.6 W on the Kintex FPGA. Both were more energy efficient than the CloudScout Myriad 2~\cite{movidiusIntelMovidiusMyriad2020} implementation (0.48 J vs. 0.63 J; Myriad 2: 346 ms, 1.8 W). Pacini et al. preserved comparable accuracy, performance, and power on the Zynq UltraScale+ MPSoC while improving on-chip memory utilization, reaching 144.8 ms at 4.51 W.

\phantomsection\label{par:yang2022}
\textbf{Yang et al. (2022)~\cite{yangAlgorithmHardwareCodesign2022}} proposed OSCAR-RT, an algorithm/hardware codesign framework for CNN-based SAR ship detection on-board satellites. This framework took target accuracy, latency, and FPGA as inputs and output candidate CNN architectures with corresponding hardware implementations. They generated these implementations using pre-built HLS CNN components and applied filter pruning and mixed-precision quantization to make them suitable for on-board computing. They evaluated this framework by generating six CNN accelerators based on MobileNetV1 \cite{howardMobileNetsEfficientConvolutional2017}, MobileNetV2 \cite{sandlerMobileNetV2InvertedResiduals2019}, and SqueezeNet \cite{iandolaSqueezeNetAlexNetlevelAccuracy2016}, trained on the SSDD dataset \cite{liShipDetectionSAR2017} and implemented on a Virtex-7 FPGA.

These OSCAR-RT accelerators achieved 93.3\%-94.6\% average precision while maintaining high performance and low power. They reached up to 366 GOP/s and consumed as little as 4.4 W. Compared to an Nvidia RTX 2080Ti GPU running the same models, OSCAR-RT reduced latency by up to 2.4x with MobileNetV2 and cut power consumption by up to 11.6x.

\phantomsection\label{par:kyriakos2022}
\textbf{Kyriakos et al. (2022)~\cite{kyriakosResourcesPowerEfficient2022}} proposed a CNN accelerator for ship detection. They designed the CNN as a lightweight model for on-board computing, comprising two convolutional layers, two pooling layers, and one FC layer. The accelerator leveraged fixed-point arithmetic in its computational units, representing NN values with 8 to 17 bits of precision. It was implemented using reusable VHDL blocks and evaluated on the Virtex-7 FPGA. The CNN model was trained on the Planet's ``Ships in Satellite Imagery'' dataset \cite{hammellShipsSatelliteImagery2018}.

This lightweight NN achieved 97.6\% accuracy in detecting ships. Notably, this accelerator enabled concurrent inference of two frames (80x80 RGB image), requiring only 0.687 ms to process a single frame. Compared with CPU and GPU execution, this accelerator achieved speedups of 6.836 and 3.205, respectively. Compared to the Jetson Nano and Myriad 2 devices, this accelerator outperformed them. However, these devices demonstrated higher performance per watt.

\phantomsection\label{par:neris2022}
\textbf{Neris et al. (2022)~\cite{nerisFPGABasedImplementationCNN2022}} proposed a CNN accelerator for a lightweight MobileNetv1~\cite{howardMobileNetsEfficientConvolutional2017} variant to detect ships and airplanes in images from the Video Imaging Demonstrator for Earth Observation (VIDEO) project~\cite{consortiumVideoImagingDemonstrator2021}. They trained the NN with ship and airplane datasets that combined MASATI~\cite{gallegoAutomaticShipClassification2018} and a subset of HRSC2016~\cite{liuHighResolutionOptical2017} (512x512 RGB), plus the UC Merced Land Use~\cite{yangBagofvisualwordsSpatialExtensions2010} and Aerial Image~\cite{xiaAIDBenchmarkData2017} datasets (256x256 RGB). They implemented two accelerator variants on a Kintex UltraScale FPGA: one with FP32 arithmetic and one with 16-bit fixed-point arithmetic.

This MobileNetv1Lite model executed about 0.05 GFLOPs for ship detection and 0.012 GFLOPs for aircraft detection. Its F1 scores were about 94.5\% and 91\%, respectively. The adopted quantization reduced LUT, FF, and block RAM (BRAM) usage while improving DSP utilization.

\phantomsection\label{par:pitonak2022}
\textbf{Pitonak et al. (2022)~\cite{pitonakCloudSatNet1FPGABasedHardwareAccelerated2022}} proposed CloudSatNet-1, a CNN accelerator that reduced CubeSat downlink by performing on-board cloud detection. They implemented it on the Zturn board, which included a Zynq Z7020 SoC, using the FINN framework~\cite{umurogluFINNFrameworkFast2017, blottFINNREndtoEndDeepLearning2018}. This framework enabled nine configurations with varying quantization (2-, 3-, and 4-bit parameter widths) and parallelism. They trained and tested the model using the Landsat 8 Cloud data (L8 biome dataset)~\cite{fogaCloudDetectionAlgorithm2017} to assess how these configurations affected False Positive Rate (FPR) and accuracy.

Excluding Snow/Ice biomes from training improved accuracy (ACC $\approx$ 92-95\%) and reduced FPR ($\approx$ 2.9-5.7\%) relative to the full dataset (ACC $\approx$ 88-90\%, FPR $\approx$ 7-10\%). Increasing parameter width further improved accuracy and reduced FPR. Parallelism also affected throughput: the non-parallel design achieved 0.9 FPS, whereas maximum layer parallelization reached 15 FPS. Power consumption ranged from 2.448 W to 2.592 W.

\phantomsection\label{par:zhang2022}
\textbf{Zhang et al. (2022)~\cite{zhangAccurateLowlatencyEfficient2022}} proposed a GNN for target recognition and an accelerator for it. They implemented this accelerator on a Zynq UltraScale+ MPSoC and optimized it, including weight and input pruning to reduce the model's memory footprint and computational requirements. The final model had 30,000 parameters and executed $2.13\times 10^{6}$ operations. This model comprised layers with GraphSAGE~\cite{hamiltonInductiveRepresentationLearning2017} operators, graph pooling layers, and Attention modules.

The authors tested this accelerator using the MSTAR dataset~\cite{sandianationallaboratoryMSTARDataset1995}, where each SAR image is 128x128 pixels. When executed at 125 MHz, the testing results demonstrated an inference time of 0.105 ms and a power consumption of 6.3 W, with an accuracy of 99.09\%. Compared to CPU and GPU implementations, these results achieved a 14.8x speedup and a 2.5x speedup, and improved energy efficiency by 62x and 39x, respectively.

\phantomsection\label{par:abderrahmane2022}
\textbf{Abderrahmane et al. (2022)~\cite{abderrahmaneSPLEATSPikingLowpower2022}} proposed SPLEAT, an SNN accelerator designed for cloud detection on board the OPS-SAT satellite on a Cyclone V \cite{evansOPSSATESANanosatellite2014,evansOPSSATOperationalConcept2016}. The authors implemented two distinct SPLEAT architectures: one using serial spike generation and the other parallel spike generation. They compared these implementations with a CNN and a Hybrid NN (HNN) implementation for the same task. These NNs were based on LeNet and trained and tested using the OPS-SAT cloud dataset. The HNN utilized conventional convolutional layers followed by spiking dense layers.

The estimated inference time and power consumption for the CNN were 25 $\mu s$ and 243.99 mW; for the HNN, 280 $\mu s$ and 295.99 mW; for the SNN utilizing serial spike generation, 1860 $\mu s$ and 41.28 mW; and for the SNN employing parallel spike generation, 160 $\mu s$ and 201.69 mW. Among the four tested ANN architectures, the SNN with serial spike generation achieved the highest accuracy of 71.92\% on the provided dataset.

\phantomsection\label{par:wang2022}
\textbf{Wang et al. (2022)~\cite{wangHardwareAccelerationImplementation2022}} proposed an accelerator for YOLOX-s~\cite{yolox2021}, a CNN for object detection trained on the COCO dataset~\cite{linMicrosoftCOCOCommon2014}. This accelerator leveraged a DSP array for convolutional layer computation and a cache system to efficiently store feature maps and weights. During design, the authors accounted for TMR's resource overhead by reserving two-thirds of available DSPs for future implementation. They implemented the accelerator on a Virtex-7 FPGA. 

This accelerator's power consumption was 14.76 W, its latency was 62.187 ms, and its throughput was 399.6 GOP/s. Additionally, the authors reported a DSP utilization rate of 85.33\% during inference, indicating efficient computation and reduced dependence on memory bandwidth.

\phantomsection\label{par:papatheofanous2022}
\textbf{Papatheofanous et al. (2022)~\cite{papatheofanousSoCFPGAAcceleration2022}} proposed the Lightweight Dilation UNet (LD-UNet), a compact model for semantic segmentation tasks. This NN was implemented on a Zynq UltraScale+ MPSoC and trained on the 95-Cloud dataset \cite{mohajeraniCloudnetCloudSegmentation2020} for cloud detection. The implementation method relied on the AMD DPU, Vitis AI, and HLS to accelerate the algorithm.

Quantization with Vitis AI resulted in degradation in accuracy, IoU, recall, and precision. Nevertheless, the FPGA implementation achieved a 1.64x speedup over the non-accelerated implementation. Additionally, according to the Vivado power analysis report, it consumed 14.111 W.

\phantomsection\label{par:zhao2023}
\textbf{Zhao et al. (2023)~\cite{zhaoHardwareAccelerationSatellite2023}} proposed a lightweight NN for onboard spacecraft object detection. They implemented it on a Kria KV260 board with a Zynq UltraScale+ MPSoC, using AMD DPU and Vitis AI. This model is a modified YOLOv4~\cite{bochkovskiyYOLOv4OptimalSpeed2020} that adopts MobileNetv3~\cite{howardSearchingMobileNetV32019} as the backbone feature extraction network, together with channel pruning and post-training quantization (PTQ). They evaluated this implementation on the DIOR dataset \cite{liObjectDetectionOptical2020}.

This evaluation showed a 91.11\% reduction in parameter size relative to the original YOLOv4 model, at the cost of 1.86\% mAP. Compared with an AMD R7-4800H CPU, it reduced power consumption by 81.91\% and increased FPS by 317.88\%. Compared with an NVIDIA RTX 2060 GPU, it reduced power consumption by 91.41\% and increased FPS by 8.50\%.

\phantomsection\label{par:mazouz2024}
\textbf{Mazouz and Nguyen (2024)~\cite{mazouzOnlineContinualStreaming2024}} proposed a NN accelerator incorporating online continuous learning (OCL), a method inspired by experience replay (ER) literature~\cite{rolnickExperienceReplayContinual2019}. This method addressed the challenges posed by variations in deployment environments. The authors used a custom dataset comprising images from the AAReST telescope~\cite{underwoodUsingCubeSatMicrosatellite2015} to train a YOLOv3 model~\cite{redmonYOLOv3IncrementalImprovement2018} for object detection. This dataset was divided into old and new categories, enabling experimentation with OCL effectiveness by varying the proportions of new and old data within input batches of variable sizes. This accelerator was implemented on a Zynq 7100 board using their CNN FPGA compiler~\cite{mazouzAutomatedCNNBackpropagation2021,mazouzAdaptiveHardwareReconfiguration2019}, with training performed via an on-board backpropagation pipeline and the proposed OCL methodology.

An implementation that utilized 100\% of available DSPs and 86\% of LUTs enabled a CNN to adapt to new environments using OCL within 3.96 minutes. However, achieving optimal average learning and forgetting values required additional iterations, with the best results obtained after five iterations (8.40 minutes). The accelerator achieved a processing rate of 90 FPS, which is sufficient for real-time benchmarking input processing.

\phantomsection\label{par:shao2024}
\textbf{Shao et al. (2024)~\cite{shaoConfigurableAcceleratorCNNBased2024}} proposed a CNN accelerator controlled with microinstructions using very long instruction words. This accelerator incorporated one or more systolic arrays that leveraged the DSP units available on the FPGA. These units efficiently performed multiply-accumulate operations and computed various CNN layer types. Furthermore, this accelerator employed 16- and 8-bit fixed-point data quantization to reduce the storage and computational load. The authors implemented this system in a Virtex-7 FPGA and evaluated it with two distinct configurations: one with a single 16-bit systolic array and another with two 8-bit systolic arrays. They then tested these implementations by inferring YOLOv3-Tiny and ResNet-18 trained on NWPU VHR-10 data~\cite{bianNWPUVHR102023}.

For a single-systolic array implementation with 16-bit quantization, the system achieved 51 FPS, 153.14 GOP/s, and 6.93 W of power consumption while executing the ResNet-18 model. For a two-systolic-array implementation with 8-bit quantization, the system achieved 102 FPS, 301.52 GOP/s, and 10.68 W of power consumption when executing the YOLOv3-Tiny model.

\phantomsection\label{par:castelino2024}
\textbf{Castelino et al. (2024)~\cite{castelinoEnergyEfficientArtefactDetection2024}} implemented a convolutional autoencoder (CAE) on the ZCU104 board to detect artifacts in HSI images using AMD DPU and Vitis AI. This CAE was trained and tested on a combined dataset of the Indian Pines, Salinas Valley, Kennedy Space Center, and University of Pavia HSI datasets. Pruning and quantization optimized this NN for FPGA inference.

The CAE achieved an average inference time of 4 ms per 144x144 HSI image, outperforming the Jetson Xavier NX GPU by factors of 4.7x (16-bit floating-point) and 2.6x (8-bit integer). Furthermore, the FPGA implementation demonstrated superior energy efficiency, consuming 21.52 mJ and outperforming the Jetson Xavier NX GPU by 7.2x and 3.6x.

\phantomsection\label{par:zhang2024}
\textbf{Zhang et al. (2024)~\cite{zhangEnergyefficientDehazingNeural2024}} proposed $E^2AOD$-Net, a real-time image dehazing NN, and an accelerator for it. They adapted $E^2AOD$-Net from AOD-Net \cite{liAODNetAllinOneDehazing2017} and leveraged several optimizations to achieve real-time performance on a PYNQ Z2 board. These optimizations included quantization, pruning, task parallelization, and pipelining. This NN was trained with the RESIDE~\cite{liBenchmarkingSingleImageDehazing2019} dataset and evaluated on D-HAZY~\cite{ancutiDHAZYDatasetEvaluate2016}, NH-HAZE~\cite{ancutiNHHAZEImageDehazing2020}, SOTS~\cite{liBenchmarkingSingleImageDehazing2019}, HSTS~\cite{liBenchmarkingSingleImageDehazing2019}, and FRIDA~\cite{tarelImprovedVisibilityRoad2010} datasets.

Compared to AOD-Net, $E^2AOD$-Net achieved higher effectiveness. Regarding implementation efficiency, this NN required 26 ms to infer one image and consumed 2.491 W. Notably, compared to the Intel Core i9-13900HX CPU, this implementation demonstrated an execution speedup of 72.885 times and a power consumption reduction of 95.5\%. Furthermore, compared to the NVIDIA GeForce RTX 4060 GPU, this implementation, although 2.6 times slower, exhibited a substantial power efficiency gain, consuming 98.1\% less power.

\phantomsection\label{par:cratere2024}
\textbf{Cratere et al. (2024)~\cite{cratereEfficientFPGAacceleratedConvolutional2024}} implemented four CNNs on the Avnet Ultra96-V2 board for cloud detection. These NNs were Pixel-Net, Patch-Net, Scene-Net, and a U-Net-based network. They were implemented using AMD DPU and Vitis AI, leveraging quantization and pruning to introduce sparsity and reduce the memory footprint. Data from Sentinel-2 Level-2A (L2A) products \cite{esaSentinel2Products2025} was used to train and test these NNs. Consequently, this method reduced the networks' parameters and operations by up to 98.6\% and 90.7\%, respectively.

The number of false positives remained below 1\%, and the accuracy loss was only 0.1-0.3\% after pruning and 0.1-0.6\% after quantization. Moreover, Scene-Net and U-Net demonstrated the lowest inference times for processing a 256x256 image, at 17.5 ms and 26.7 ms, respectively.

\phantomsection\label{par:kimOn2024}
\textbf{Kim et al. (2024)~\cite{kimOnOrbitAICloud2024}} proposed a 3-stage cloud detection methodology. Initially, an image uniformity check filtered out entirely clouded or clear-sky images, and subsequently, a NN classifier, TriCloudNet, based on SqueezeNet \cite{iandolaSqueezeNetAlexNetlevelAccuracy2016}, filtered out extreme cases. Concurrently, a U-Net-based segmentation NN classified each pixel as either cloud or non-cloud. The authors accelerated these NNs on a Zynq 7000 SoC using HLS. This FPGA is identical to the on-board computer in the 6U-class A-HiREV nanosatellite \cite{zelekeNewStrategySatellite2023}. Moreover, they optimized the U-Net for nanosatellites by pruning it. These NNs were trained and tested on a combined dataset comprising 96 x 96 RGB images from the SPARCS \cite{karakiziDetailedLandCover2018} and Landsat 8 CCA \cite{irishCharacterizationLandsat7ETM2006} datasets.

This implementation achieved a 6.21x speedup over the software-only implementation, outperforming Nvidia Jetson Nano-based acceleration. Notably, this work showed that the multi-stage filtering method reduced processing time and power consumption by approximately 48\% and can reduce downlink data volume by 40-50\%.

\subsection{Spacecraft Perception, Autonomy, and Navigation}
Most implementations discussed in this subsection used off-board, supervised training. However, exceptions include Gankidi and Thangavelautham (2017)~\cite{gankidiFPGAArchitectureDeep2017}, who executed an on-board reinforcement learning method (Q-learning), and Lent (2020)~\cite{lentEvaluatingCognitiveNetwork2020}, who performed on-board unsupervised training. Table~\ref{tab:autonomy_and_navigation} summarizes key methodological differences among the works in this subsection.

\begin{table*}[htbp]
  \centering
  \small
  \caption{Distinctive characteristics of each spacecraft perception, autonomy, and navigation system reviewed in this section}
  \label{tab:autonomy_and_navigation}
  \footnotesize
  \renewcommand{\arraystretch}{1.2}
  \begin{tabular}{llllll}
    \toprule
    \textbf{Key.} & \makecell[tl]{\textbf{Data}\\\textbf{Origin}} & \makecell[tl]{\textbf{Task}\\\textbf{Type}} & \makecell[tl]{\textbf{NN}\\\textbf{Architecture}} & \makecell[tl]{\textbf{FPGA}\\\textbf{Family}}   & \makecell[tl]{\textbf{Hardware}\\\textbf{Design}} \\
    \midrule
    \hyperref[par:gankidi2017]{\makecell[tl]{\textbf{Gankidi and}\\\textbf{Thangavelautham (2017)~\cite{gankidiFPGAArchitectureDeep2017}}}} & In-Situ & Classification & MLP & Virtex 7 & -- \\
    \textbf{\hyperref[par:lent2020]{Lent (2020)~\cite{lentEvaluatingCognitiveNetwork2020}}} & In-Situ & Classification & SNN & Zynq 7000 SoC & HLS \\
    \hyperref[par:cosmas2020]{\makecell[tl]{\textbf{Cosmas and}\\\textbf{Kenichi (2020)~\cite{cosmasUtilizationFPGAOnboard2020}}}} & Remote & \makecell[tl]{Keypoint\\Detection} & CNN & \makecell[tl]{Zynq UltraScale+\\ MPSoC} & AMD DPU \\
    \textbf{\hyperref[par:jiang2023]{Jiang and Sha (2023)~\cite{jiangRFFingerprintingIdentification2023}}} & In-Situ & Classification & SNN & Virtex 7 & -- \\
    \textbf{\hyperref[par:ekblad2023]{Ekblad et al. (2023)~\cite{ekbladResourceconstrainedFPGADesign2023}}} & Remote & \makecell[tl]{Object\\Detection} & CNN &  \makecell[tl]{Zynq UltraScale+\\ MPSoC} & AMD DPU \\
    \hyperref[par:carmeli2023]{\makecell[tl]{\textbf{Carmeli and}\\\textbf{Ben-Moshe (2023)~\cite{carmeliAIBasedRealTimeStar2023}}}} & Remote & Classification & SOM & Cyclone V & -- \\
    \textbf{\hyperref[par:guo2024]{Guo et al. (2024)~\cite{guoOverlayAcceleratorDeepLab2024}}} & Remote & Segmentation & CNN & Virtex 7 & Verilog HDL \\
    \textbf{\hyperref[par:kim2024]{Kim et al. (2024)~\cite{kimFPGAAcceleratedCNNParallelized2024}}} & In-Situ & Classification & CNN & Zynq 7000 SoC & HLS \\
    \textbf{\hyperref[par:li2025]{Li et al. (2024)~\cite{liFPGABasedLowBitLightweight2025}}} & In-Situ & \makecell[tl]{Depth\\Estimation} & CNN &  \makecell[tl]{Zynq UltraScale+\\ MPSoC} & FINN \\
    \bottomrule
  \end{tabular}
\end{table*}

\phantomsection\label{par:gankidi2017}
\textbf{Gankidi and Thangavelautham (2017)~\cite{gankidiFPGAArchitectureDeep2017}} developed a NN incorporating Q-learning to enhance spacecraft autonomy and decision-making capabilities. They implemented this work on a Virtex-7 FPGA, utilizing two distinct approaches. The first method employed a single-neuron accelerator, while the second utilized an MLP accelerator. These methods leveraged floating-point or fixed-point arithmetic operations. While fixed-point arithmetic reduced accuracy, it also lowered power consumption. Its impact was evaluated in two distinct scenarios, comparing both accelerators under varying conditions. The MLP was implemented with 11 neurons for the simplest scenario and 25 for the most complex one.

The single-neuron accelerator achieved up to 95 times the speed of an Intel i5 CPU. Similarly, the MLP accelerator achieved a maximum 43x increase in efficiency compared to this CPU. Regarding power consumption, this latter implementation consumed 5.6 W with fixed-point arithmetic in the simplest scenario and up to 9.7 W with floating-point arithmetic in the more complex scenario.

\phantomsection\label{par:lent2020}
\textbf{Lent (2020)~\cite{lentEvaluatingCognitiveNetwork2020}} proposed an implementation of the Cognitive Network Controller (CNC) \cite{lentCognitiveNetworkController2018} to address routing challenges in Delay-tolerant networking~\cite{lentNeuromorphicArchitectureDisruption2019} on-board satellites. This implementation leveraged an SNN to decide on satellite outbound connections and to store information about its decision-making performance. The CNC employed an SNN with Spike-Timing Dependent Plasticity (STDP)~\cite{stdp}, meaning it can be applied to new scenarios without prior training and will adapt and become more proficient over time. The SNN architecture in this CNC utilized a Leaky-Integrate-and-Fire (LIF) neuron model and was implemented on a Zynq 7020 SoC with 256 neurons. Additionally, the author tested this implementation on a 5-node network testbed.

The results demonstrated the advantage of this implementation over a CPU-only solution, showing lower response time and higher throughput than Contact Graph Routing, which served as the reference. Notably, leveraging the FPGA fabric to accelerate the CNC SNN inference, this system executed the SNN inference 32 times faster than its software-only counterpart.

\phantomsection\label{par:cosmas2020}
\textbf{Cosmas and Kenichi (2020)~\cite{cosmasUtilizationFPGAOnboard2020}} presented CNN acceleration for spacecraft pose estimation, leveraging Vitis AI~\cite{amdVitisAIUser2025} on an Ultra96-V2 board featuring an UltraScale+ MPSoC. Their implementation had three primary stages: spacecraft detection, image cropping, and heatmap-based key points detection. The authors employed YOLOv3~\cite{redmonYOLOv3IncrementalImprovement2018} for spacecraft detection and image cropping, with the resultant output being resized to 128x128 pixels before input into a ResNet34-U-Net model for key points detection. They optimized these networks through pruning and quantization with Vitis AI and evaluated them on three distinct AMD DPU architectures. However, due to hardware constraints, they only implemented one instance of these DPUs on the FPGA. The proposed system was trained and evaluated using the SPEED dataset~\cite{kisantalSatellitePoseEstimation2020}.

The Root Mean Square Error (RMSE) difference between the FPGA and PC implementations was less than 0.55, with average RMSEs of 1.913 for the FPGA and 1.382 for the PC across 100 images. Notably, the FPGA implementation exhibited on-chip power consumption of 3.4-3.9 W, while the board consumed 8.6 W during inference.

\phantomsection\label{par:jiang2023}
\textbf{Jiang and Sha (2023)~\cite{jiangRFFingerprintingIdentification2023}} proposed a radio frequency fingerprinting identification method leveraging SNNs. They designed an SNN with LIF neurons and implemented it on a Virtex-7 FPGA. They trained this NN using Spatiotemporal backpropagation (STBP) \cite{wuSpatiotemporalBackpropagationTraining2018} on a dataset comprised of 56,000 MATLAB simulation samples obtained at different signal-to-noise ratios (SNRs).

At an SNR of 25 dB, the model achieved an accuracy of 95.26\%, inferred a sample in 27.14 ms, and consumed 1.78 W.

\phantomsection\label{par:ekblad2023}
\textbf{Ekblad et al. (2023)~\cite{ekbladResourceconstrainedFPGADesign2023}} implemented a YOLOv4-based model for satellite component detection. They replaced the Mish activation function~\cite{misraMishSelfRegularized2020} with leaky ReLU~\cite{maasRectifierNonlinearitiesImprove2013}, and reduced the max pooling kernel size to 8x8 of the original YOLOv4~\cite{bochkovskiyYOLOv4OptimalSpeed2020} to accommodate Vitis AI. Furthermore, they evaluated the NN on a Kria KV260, which features a Zynq UltraScale+ MPSoC. The dataset used in this evaluation consisted of images captured at the Florida Institute of Technology's ORION facility utilizing the Kinematics Simulator.

This implementation outperformed a previous work~\cite{mahendrakarAutonomousRendezvousNoncooperative2023}, where the authors deployed the YOLOv5 model~\cite{jocherYOLOv5Ultralytics2020} on an Intel NCS2 and Raspberry Pi, achieving 3.8 FPS compared to 2 FPS. Nevertheless, its mAP decreased by 2.9\% relative to the original model due to quantization and FPGA implementation.

\phantomsection\label{par:carmeli2023}
\textbf{Carmeli and Ben-Moshe (2023)~\cite{carmeliAIBasedRealTimeStar2023}} proposed a SOM accelerator for identifying star patterns on-board satellites as part of an end-to-end environment that encompasses image capture, processing, and spacecraft position identification. This system leveraged a DE1-SoC board, a high-sensitivity VIS camera, a display showing the captured image, and a graphical user interface (GUI) application for testing and displaying results. Upon initialization by the ARM processor, the system filtered the captured image and executed the main algorithm while interfacing with the Cyclone-V FPGA. The latter controlled the camera for image capture, processed the images using the SOM, and interfaced with the GUI application. The SOM received a feature vector derived from the positions of stars in the image as input and compared it to data stored in the Almanac.

The proposed accelerator enabled efficient identification of star patterns while minimizing memory utilization. Notably, the SOM executed in approximately 870 $\mu s$ and required 249 KB of memory, with a confidence level of roughly 98\% in its ability to accurately identify star patterns.

\phantomsection\label{par:guo2024}
\textbf{Guo et al. (2024)~\cite{guoOverlayAcceleratorDeepLab2024}} proposed a CNN accelerator for semantic spacecraft component images (SCIs) segmentation. Compared to other CNN accelerators for FPGAs, this overlay approach allowed the execution of various CNN models without changing the hardware implementation. To achieve this implementation, the authors worked on a COD (Control, Operation, and Data Transfer) Instruction Set Architecture (ISA), a compiler for this ISA, and a hardware accelerator design. They implemented this accelerator on a Virtex-7 FPGA and tested using six CNN models trained with the Spacecraft Dataset~\cite{dungSpacecraftDatasetDetection2021} and the SCIs Dataset~\cite{guoSCIsDataset2018}. These CNN models incorporated a combination of a backbone network (VGG16~\cite{simonyanVeryDeepConvolutional2015}, ResNet18~\cite{heDeepResidualLearning2015}, or SqueezeNet1.1~\cite{iandolaSqueezeNetAlexNetlevelAccuracy2016}) and a head network (Deeplabv3+~\cite{chenEncoderDecoderAtrousSeparable2018} or DeepLabv3~\cite{chenRethinkingAtrousConvolution2017}).

The authors compared the results with inference on existing CNN image segmentation accelerators, other FPGA overlay solutions, and an NVIDIA RTX 2080 Ti GPU. This implementation demonstrated higher computational resource efficiency than previous image segmentation accelerators and other FPGA overlay solutions, achieving 159.48 GOP/s while using fewer hardware resources. While not outperforming the GPU, this implementation exhibited a 5.1x improvement in energy efficiency, consuming 21 W.

\phantomsection\label{par:kim2024}
\textbf{Kim et al. (2024)~\cite{kimFPGAAcceleratedCNNParallelized2024}} proposed a dueling DQN-based reinforcement learning algorithm. They designed this algorithm for real-time routing in LEO satellite networks and accelerated it on a PYNQ-Z2 board. A key component of this algorithm is a CNN implemented on the FPGA using HLS. This CNN comprised convolutional and ReLU layers.

The FPGA-based acceleration outperformed CPU execution by a factor of 3.10. Specifically, the average execution time of this implementation was 0.2991 s.

\phantomsection\label{par:li2025}
\textbf{Li et al. (2024)~\cite{liFPGABasedLowBitLightweight2025}} proposed $L^3FNet$, a low-bit, lightweight light field depth estimation network. This network was implemented on a ZCU104 board using FINN. The authors employed disparity partitioning preprocessing, two-dimensional network architecture optimization, pruning, and quantization to adapt the network for efficient execution. The quantization method utilized 8-bit values in the input layer and 4-bit values for the parameters stored in on-chip memory. They trained and evaluated the network using the 4D LF Benchmark \cite{honauerDatasetEvaluationMethodology2017} dataset.

The $L^3FNet$ system achieved an inference time of 0.272 ms per image, which is 1210 times faster than its floating-point counterpart on an Intel i5-9600K CPU, 40 times faster than the NVIDIA 2060 SUPER GPU, and 33 times faster than the NVIDIA RTX 3090 GPU. Furthermore, the developed system exhibited a power consumption of 9.493 W, which was 6.5 times lower than that of an Intel i5-9600k CPU, 18 times lower than that of an NVIDIA 2060 super GPU, and 35 times lower than that of an NVIDIA RTX 3090 GPU.

\subsection{Anomaly Detection and Undefined Use Cases}
This subsection covers anomaly-detection accelerators together with undefined use cases: FPGA accelerators intended for onboard space computing but evaluated without a mission-specific dataset or benchmark. All works in this subsection used off-board training, and most relied on supervised learning. The exception is Ma et al. (2019)~\cite{maLightweightHyperspectralImage2019}, who employed an unsupervised approach. Notably, Kesuma et al. (2019)~\cite{kesumaArtificialIntelligenceImplementation2019}, Ma et al. (2019)~\cite{maLightweightHyperspectralImage2019}, and Pacini et al. (2023)~\cite{rnn_pacini} implemented lightweight architectures with smaller input layers and lower parameter counts. Table~\ref{tab:anomaly} summarizes additional key algorithmic and hardware distinctions.

\begin{table*}[htbp]
  \centering
  \small
  \caption{Distinctive characteristics of the anomaly-detection and undefined-use-case systems reviewed in this subsection}
  \label{tab:anomaly}
  \footnotesize
  \renewcommand{\arraystretch}{1.2}
  \begin{tabular}{llllll}
    \toprule
    \textbf{Key.} & \makecell[tl]{\textbf{Data}\\\textbf{Origin}} & \makecell[tl]{\textbf{Task}\\\textbf{Type}} & \makecell[tl]{\textbf{NN}\\\textbf{Architecture}} & \makecell[tl]{\textbf{FPGA}\\\textbf{Family}}   & \makecell[tl]{\textbf{Hardware}\\\textbf{Design}} \\
    \midrule
    \textbf{\hyperref[par:kesuma2019]{Kesuma et al. (2019)~\cite{kesumaArtificialIntelligenceImplementation2019}}} & In-Situ & Classification & MLP & Virtex-5QV & LEON3 \\
    \textbf{\hyperref[par:ma2019]{Ma et al. (2019)~\cite{maLightweightHyperspectralImage2019}}} & Remote  & Segmentation & \makecell[tl]{Fully-Connected\\ Autoencoder} & \makecell[tl]{Zynq UltraScale+\\ MPSoC} & HLS \\
    \textbf{\hyperref[par:perryman2023]{Perryman et al. (2023)~\cite{perrymanEvaluationXilinxVersal2023}}} & -- & Classification & CNN & Versal & HLS \\
    \textbf{\hyperref[par:coca2023]{Coca and Datcu (2023)~\cite{cocaFPGAAcceleratorMetaRecognition2023}}} & Remote & Classification & CNN & \makecell[tl]{Zynq UltraScale+\\ MPSoC} & AMD DPU \\
    \textbf{\hyperref[par:pacini2023]{Pacini et al. (2023)~\cite{rnn_pacini}}} & In-Situ & Classification & RNN & Kintex UltraScale & HDL \\
    \textbf{\hyperref[par:benelli2024]{Benelli et al. (2024)~\cite{gpu_at_sat}}} & -- & Classification & CNN & \makecell[tl]{Zynq UltraScale+\\ MPSoC} & GPU@SAT \\
    \bottomrule
  \end{tabular}
\end{table*}

\phantomsection\label{par:kesuma2019}
\textbf{Kesuma et al. (2019)~\cite{kesumaArtificialIntelligenceImplementation2019}} proposed the implementation of an AI Kit capable of detecting faults and receiving commands from the astronauts. This kit implemented electronic circuits that preprocessed data from a microphone and various sensors on board a spacecraft before sending it to the central computer for processing. This AI Kit communicated with the primary computer system through a UART interface. This computer system comprised a LEON3 processor implemented on a Virtex-5QV FPGA \cite{ug190Virtex5FPGAUser2009}, a radiation-hardened FPGA, and ran a feed-forward NN under RTEMS.

The authors trained and evaluated the ANN's performance in detecting 3 distinct voice commands and anomalies in sensor data. The results demonstrated that this system could accurately classify voice commands with 99.47\% accuracy in 31.74 ms after training for 97.90 seconds. After 28.47 s of training, the system achieved 83.3\% anomaly-detection accuracy and a latency of 14.32 ms.

\phantomsection\label{par:ma2019}
\textbf{Ma et al. (2019)~\cite{maLightweightHyperspectralImage2019}} proposed the Pruning-Quantization-Anomaly Detector (P-Q-AD) accelerator for anomaly detection in Earth HSI. This accelerator combined network pruning and quantization within an autoencoder NN. The authors evaluated four implementations: one using floating-point arithmetic, one using floating-point arithmetic with weight pruning, one using 32-bit fixed-weight quantization and pruning, and one using custom weight quantization and pruning. They determined the arithmetic precision of this quantization by solving a multi-objective optimization problem. These architectures were executed on a Zynq UltraScale+ FPGA and compared with Local Reed-Xiaoli~\cite{reedAdaptiveMultiplebandCFAR1990} and collaborative representation-based detectors~\cite{liCollaborativeRepresentationHyperspectral2015} running on two Intel Xeon CPUs. The implementations were trained and evaluated on three HSI datasets from Louisiana, San Diego airport in California, and Los Angeles in the USA, acquired by the Airborne Visible Infrared Imaging Spectrometer (AVIRIS)~\cite{hamlinImagingSpectrometerScience2010}.

P-Q-AD outperformed traditional methods by two orders of magnitude and floating-point implementations by one order of magnitude, while maintaining an Area Under the Curve (AUC) difference of less than 0.01. Specifically, this implementation yielded AUC values of 0.9973, 0.9483, and 0.9869 for the Louisiana, San Diego, and Los Angeles datasets.

\phantomsection\label{par:perryman2023}
\textbf{Perryman et al. (2023)~\cite{perrymanEvaluationXilinxVersal2023}} analyzed the performance of the AMD Versal Adaptive SoC~\cite{ouellette2022system} for CNN computation in potential image classification algorithms. They computed MobileNetV1 \cite{howardMobileNetsEfficientConvolutional2017}, ResNet-50~\cite{heDeepResidualLearning2015}, and GoogLeNet~\cite{szegedyGoingDeeperConvolutions2015} on different components of this platform: the ARM Cortex-R5F CPU, the ARM Cortex-A72 CPU, the FPGA fabric, and the AI engines. Each of these implementations was compared with respect to computing speed, power consumption, and FPGA fabric resource consumption. Furthermore, the performance of these implementations was compared with the same applications running on the AMD Zynq 7-series SoC.

The Versal FPGA fabric achieved the highest computing speed while being the most power-consuming implementation. When executing the NN applications, the Versal FPGA achieved a 36.73x speedup over its ARM Cortex-R5F CPU counterpart while consuming 33.04 W. Regarding resource utilization, the Versal AI engines implementation almost entirely avoided utilizing FPGA fabric resources, and the FPGA fabric implementation used only a small percentage of available resources. Furthermore, the results from the Zynq 7-series SoC implementations were identical to those obtained on the Versal platform.

\phantomsection\label{par:coca2023}
\textbf{Coca and Datcu (2023)~\cite{cocaFPGAAcceleratorMetaRecognition2023}} implemented a ResNet model~\cite{koonceResNet502021} on an FPGA to detect natural anomalies, such as wildfires, in multispectral imagery. This model was trained and tested on three datasets: the BigEarthNet~\cite{sumbulBigearthnetLargeScaleBenchmark2019,sumbulBigEarthNetMMLargeScaleMultimodal2021} and two Sentinel-2 products~\cite{gattiSentinel2ProductsSpecification2015}, Zamora and Bordeaux. The authors initially trained the model on a machine with an Intel Xeon CPU and a Tesla K80 GPU, and then deployed it on a ZCU102 board using the AMD DPU and Vitis AI. The board's CPU processes the CNN's input and output, while the FPGA executes the CNN.

Quantization reduced the accuracy of the original FP32 model from 81.1\% to 73.7\% on the BigEarthNet dataset. However, this implementation demonstrated several advantages over the Tesla K80 GPU, including a 0.7 s inference time versus 3.128 s, an 85.677 FPS versus 19.861 FPS, and 30 W of power consumption versus 135 W. Compared with other platforms such as the RPi-Movidius system \cite{delrossoOnboardVolcanicEruption2021} and Ma et al. (2019) \cite{maLightweightHyperspectralImage2019}, this implementation achieved 2.24x and 4.16x speedups, respectively, and reduced hardware resource consumption.

\phantomsection\label{par:pacini2023}
\textbf{Pacini et al. (2023)~\cite{rnn_pacini}} integrated Recurrent Neural Networks into FPG-AI~\cite{fpg_ai}, an end-to-end NN acceleration framework, to accelerate two models based on Gated Recurrent Units. These models were evaluated in a space-oriented Fault Detection, Isolation, and Recovery case study to forecast housekeeping signals from the ESA Mars Express mission. The authors employed uniform-symmetric quantization to convert FP32 values into their integer equivalents. This quantization, combined with a custom hardware architecture, enabled the implementation of these models on a radiation-tolerant Kintex UltraScale XQRKU060 FPGA.

Compared to a Jetson Nano implementation, the FPGA-based approach consumed one order of magnitude less power (0.728 W and 0.874 W versus 2.522--2.918 W and 2.549--3.121 W) and reduced inference latency by two orders of magnitude (4.56 $\mu$s and 15.87 $\mu$s versus 833.72--510.74 $\mu$s and 1409.25--863.26 $\mu$s). Consequently, these optimizations resulted in a two-order-of-magnitude reduction in energy per inference.

\phantomsection\label{par:benelli2024}
\textbf{Benelli et al. (2024)~\cite{gpu_at_sat}} presented GPU@SAT devKit, a soft-GPU implemented on a Zynq UltraScale+ MPSoC. This architecture enables AI workloads to run on board spacecraft in a GPU. The authors evaluated it by executing the Cifar-10 NN from the GPU4S benchmark and measuring execution times for various kernels. To further demonstrate its suitability for space applications, they also synthesized it for radiation-tolerant FPGAs in a separate work~\cite{gpu_at_sat_rt_fpga}.

At 250 MHz, it achieved 35 FPS.

\section{Observations, Discussion, and Future Directions}
\label{sec:Section_4}
After analyzing individual studies, we broaden our perspective to identify prevailing trends, existing gaps, and potential directions for future research. Throughout the literature, there were two main approaches to this field. Researchers either approached FPGA implementation from the NN perspective or NN acceleration from the FPGA perspective. These works aimed to achieve lower inference time and power consumption than COTS approaches.

\subsection{Task Type and Neural Network Architecture}
\begin{wrapfigure}{r}{0.5\textwidth}
\centering
\includegraphics[width=\linewidth]{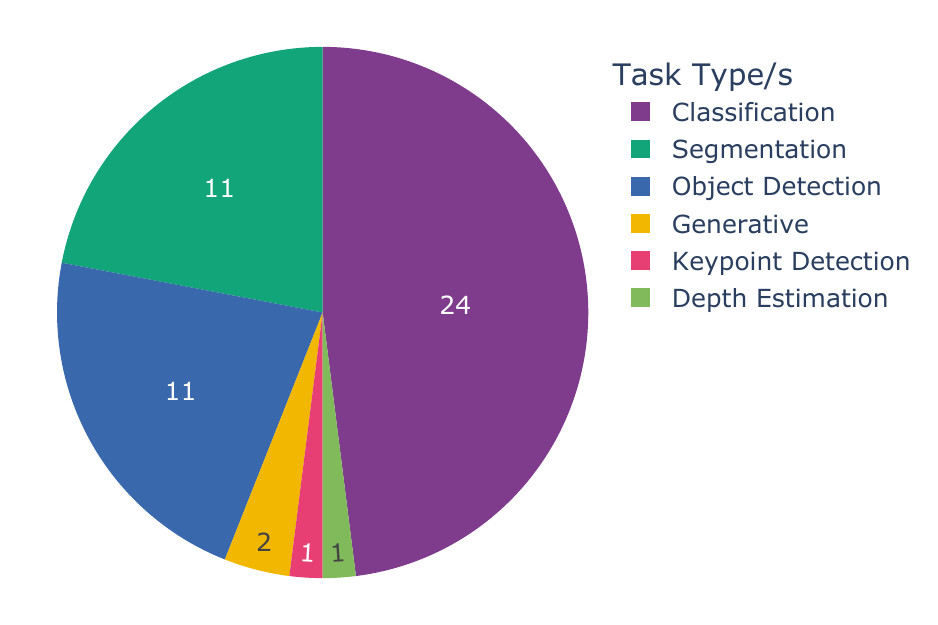}
\caption{\label{fig:nn_architectures}Task types found in the literature}
\Description[Neural Network types used in the surveyed literature]{Neural Network types used in the surveyed literature. The pie chart shows that most works employ classification networks, followed by segmentation and object detection networks. Other task types are less common, and transformers are not used in any of the works.}
\end{wrapfigure}
Figure~\ref{fig:nn_architectures} shows the distribution of NN tasks reported in the literature. Classification dominates, appearing in 51\% of the surveyed work. This type of task is primarily addressed with CNNs, including custom architectures and established models such as YOLO, MobileNet, and ResNet; Pacini et al. (2023)~\cite{rnn_pacini} are the exception, accelerating GRU-based RNNs. Segmentation and object detection are each reported in 23\% of the surveyed papers. Notably, many of these NNs target EO applications, such as cloud detection~\cite{bahlLowpowerNeuralNetworks2019,reiterFPGAAccelerationQuantized2020,lemaireFPGAbasedHybridNeural2020,rapuanoFPGABasedHardwareAccelerator2021,paciniMultiCacheSystemOnChip2021,pitonakCloudSatNet1FPGABasedHardwareAccelerated2022,abderrahmaneSPLEATSPikingLowpower2022,papatheofanousSoCFPGAAcceleration2022,cratereEfficientFPGAacceleratedConvolutional2024,kimOnOrbitAICloud2024}.

Future research should broaden its scope to include diverse NN tasks and architectures, such as generative models, Transformer-based networks~\cite{transformers}, and spike-based networks~\cite{snn_survey}. Generative models can reduce downlink volume and restore degraded imagery. Autoencoders and variational autoencoders (VAEs) have already been used for on-board compression, including an in-orbit VAE that encodes image tiles into compact latents for downstream tasks~\cite{data_compression,ravaen}; the encoder can run on the spacecraft while the decoder reconstructs the data on the ground~\cite{mcsoc_paper}. GANs and diffusion models further support cloud removal, super-resolution, and data augmentation~\cite{gan_rs_survey,diffusion_rs}. Spike-based networks remain attractive for event-driven energy efficiency and unsupervised on-board learning (Section~\ref{sec:training}). Szczerek and Podobas~\cite{snn_survey} survey this design space. In addition, many other NN architectures remain unexplored. Notably, none of the reviewed literature supports five-dimensional tensor inputs or 3D convolutions, although these capabilities are important for specific space applications~\cite{3d_convolution,mcsoc_paper}.

\subsection{Training and Learning}
\label{sec:training}
The majority of NNs were trained off-board using supervised learning. These NNs were pre-trained using previously built and labeled datasets, which might not be feasible for new sensors or unexplored environments. Due to these restrictions, pre-trained NNs could not adapt effectively to new environments. Only 2 papers employed off-board unsupervised approaches~\cite{maLightweightHyperspectralImage2019,castelinoEnergyEfficientArtefactDetection2024}, 1 explored on-board supervised training~\cite{mazouzOnlineContinualStreaming2024}, and 1 combined on-board learning with unsupervised methods~\cite{lentEvaluatingCognitiveNetwork2020}. The datasets and NNs' input primarily consist of remote sensing images from satellite sensors, including RGB, SAR, and multispectral images. These inputs range from single pixels with 12 spectral bands to full 1024x1024x3 RGB images.

Developing NNs that can learn on board could enhance adaptability to dynamic environments and reduce reliance on labeled datasets. The same constraint drives unsupervised anomaly detection in spacecraft operations: Ruszczak et al. released OPSSAT-AD, a real ESA CubeSat telemetry benchmark, and argue that representative anomaly labels are inherently limited~\cite{opssat_ad}. SNNs with STDP represent one possible solution, but they are not the only option~\cite{unsupervised_learning_survey,snn_survey}.

\subsection{Size and Parameters}
The number of parameters in the surveyed NNs varies widely, from a few hundred to millions, depending on the model. This number increases with the number of neurons. The NN's memory footprint depends on the parameter count and the quantization strategy. Most NNs employ 8-bit fixed-point quantization. Nevertheless, two implementations adopted a 1-bit quantization strategy~\cite{hashimotoShipClassificationSAR2018,reiterFPGAAccelerationQuantized2020}, and a few others retained FP32 precision~\cite{chenFPGABasedImplementation2018,pitsisEfficientConvolutionalNeural2019,bahlLowpowerNeuralNetworks2019,nerisFPGABasedImplementationCNN2022,zhangAccurateLowlatencyEfficient2022,perrymanEvaluationXilinxVersal2023}. Quantization can be uniform across all layers or customized per layer~\cite{rapuanoFPGABasedHardwareAccelerator2021}. Most NNs had a memory footprint of a few megabytes or less. The largest reported NN footprint was 49.4 MB~\cite{carmeliAIBasedRealTimeStar2023,niAlgorithmHardwareCoOptimizationDeployment2023}.

The surveyed studies show that a slight degradation in NN accuracy due to quantization is common but does not impede the NN objective. Some literature minimizes this degradation by performing Quantization-Aware Training (QAT) or fine-tuning their NN on a subset of their dataset following Post-Training Quantization (PTQ). Alternatively, some reports employ PTQ exclusively since it is less computationally intensive.

BNNs and layer-wise quantization remain largely unexplored in space applications. These methods can significantly reduce memory usage and computational complexity. Their implementation in FPGAs could lead to more efficient on-board processing.

\subsection{Implementation Strategies}
\begin{wrapfigure}{r}{0.5\textwidth}
\centering
\includegraphics[width=\linewidth]{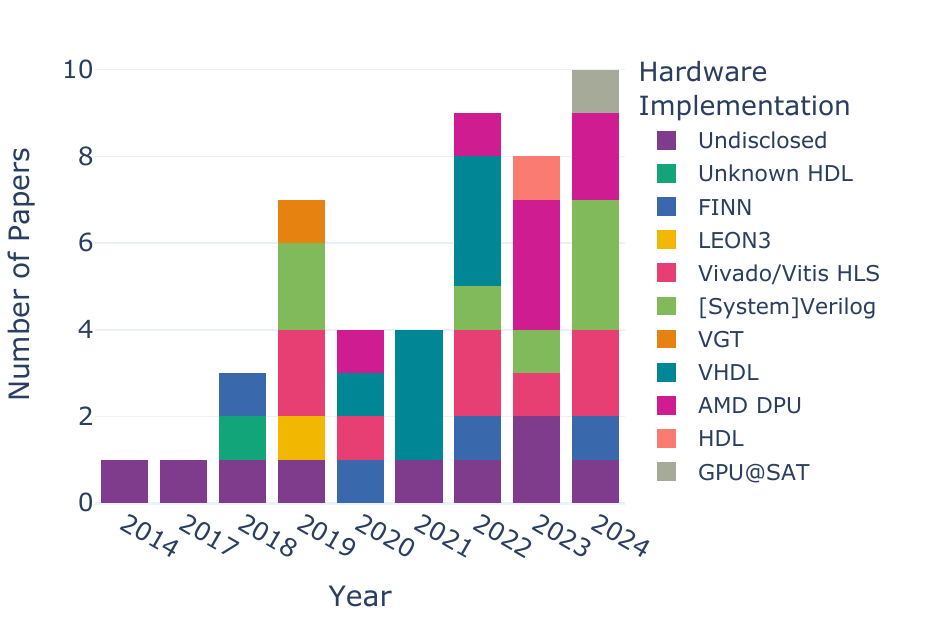}
\caption{\label{fig:hardware_implementation_by_year}Distribution of hardware implementation methods by year in the surveyed literature}
\Description[Distribution of hardware implementation methods by year in the surveyed literature]{Stacked bar chart showing the distribution of hardware implementation methods across the surveyed literature from 2014 to 2024. The chart highlights that while HDLs (such as Verilog, SystemVerilog, and VHDL) dominate overall, higher-level frameworks and particularly AMD DPU implementations have grown in recent years.}
\end{wrapfigure}
Most papers implemented their NNs on AMD FPGAs, with only 5 studies using Altera FPGAs and 1 study not specifying its FPGA. Most of these implementations leveraged HDL. However, the use of AMD's DPU is increasingly popular, with 5 implementations in the last 2 years (see Figure~\ref{fig:hardware_implementation_by_year}). Across all 47 accelerators, 24 used a time-multiplexed architecture, 16 used a dataflow architecture, 5 did not describe their logic design, and 2 used general-purpose hardware (LEON3 and GPU@SAT). Authors typically choose time-multiplexed architectures for NNs with larger parameter counts. Still, the implementation that achieved the highest OP/s used a dataflow architecture~\cite{liFPGABasedLowBitLightweight2025}. The choice between these two architectures depends on the NN model and the mission's objectives. Most of these accelerators used a clock signal at either 200 or 100 MHz, with dataflow accelerators typically operating at lower frequencies than time-multiplexed accelerators. A few accelerators employed two distinct clock signals~\cite{sabogalReCoNReconfigurableCNN2019,rapuanoFPGABasedHardwareAccelerator2021,sabogalReconfigurableFrameworkResilient2021,paciniMultiCacheSystemOnChip2021,castelinoEnergyEfficientArtefactDetection2024}, which served as the clock for separate hardware components, such as accelerator interconnects and DSPs.

The AMD DPU gained popularity since NN deployment is easier for non-hardware experts with Vitis AI. On the other hand, tools such as VGT and FINN have not gained as much popularity. This limited adoption may indicate a preference for more general-purpose tools over highly specialized solutions or a lack of tool maturity. Future research could focus on developing more tools that synthesize NNs into RTL and ease their implementation on FPGAs.

\subsection{Performance and Power-consumption Trends}
\begin{figure}[htbp]
\centering
\begin{subfigure}[b]{0.49\textwidth}
\centering
\includegraphics[width=\linewidth]{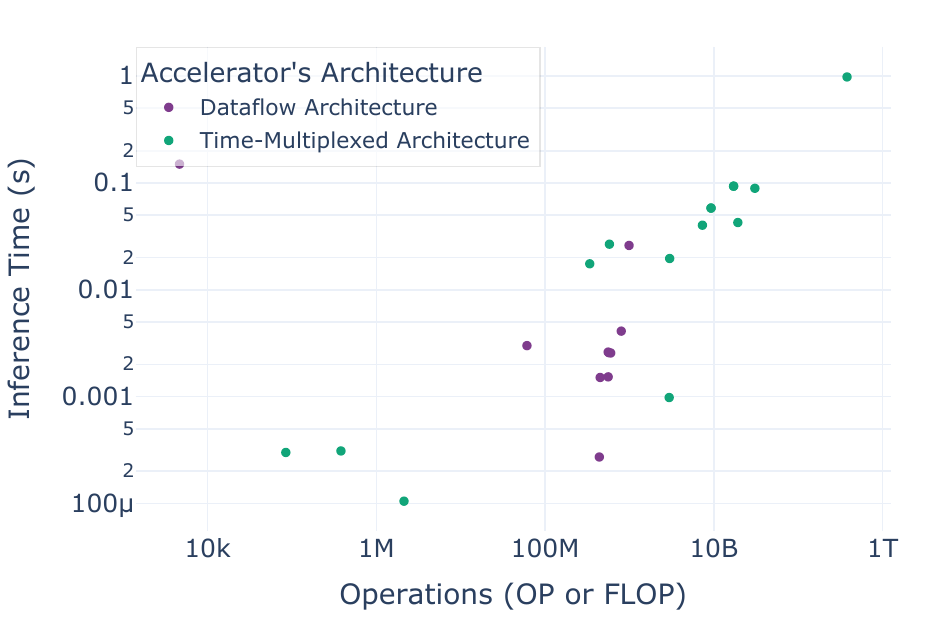}
\caption{Number of operations vs. inference time}
\label{fig:operations_vs_inference_time}
\end{subfigure}
\hfill
\begin{subfigure}[b]{0.49\textwidth}
\centering
\includegraphics[width=\linewidth]{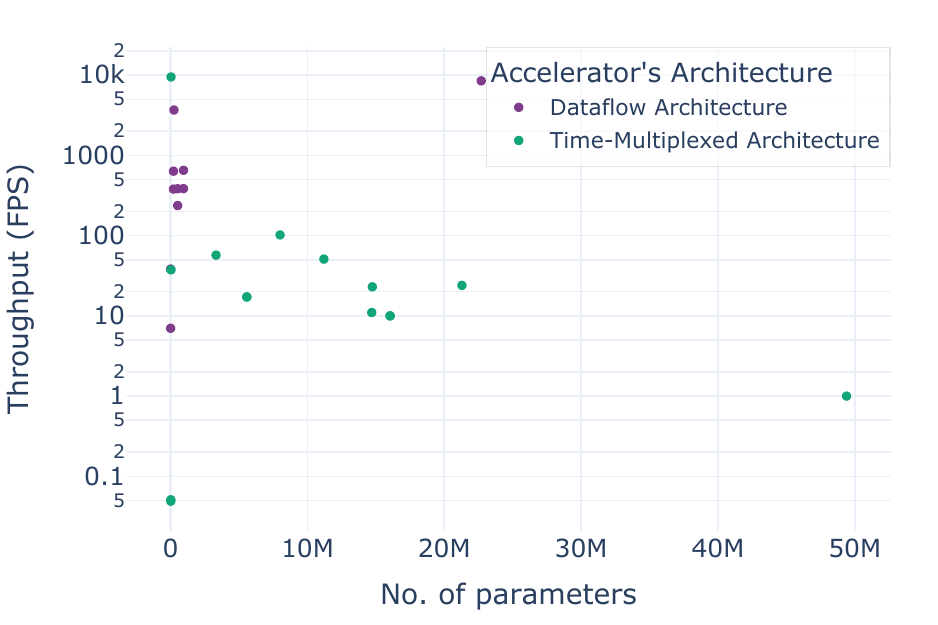}
\caption{Number of parameters vs. throughput}
\label{fig:parameters_vs_throughput}
\end{subfigure}
\caption{Influence of operations and parameters on inference time and throughput}
\label{fig:operations_parameters}
\Description[Influence of operations and parameters on inference time and throughput]{Influence of operations and parameters on inference time and throughput. The first figure shows the number of operations vs. inference time, while the second figure shows the number of parameters vs. throughput. In the first figure, we see an increase in inference time with the number of operations, while in the second figure, we see a decrease in throughput with the number of parameters.}
\end{figure}

\begin{figure}[htbp]
\centering
\begin{subfigure}[b]{0.49\textwidth}
\centering
\includegraphics[width=\linewidth]{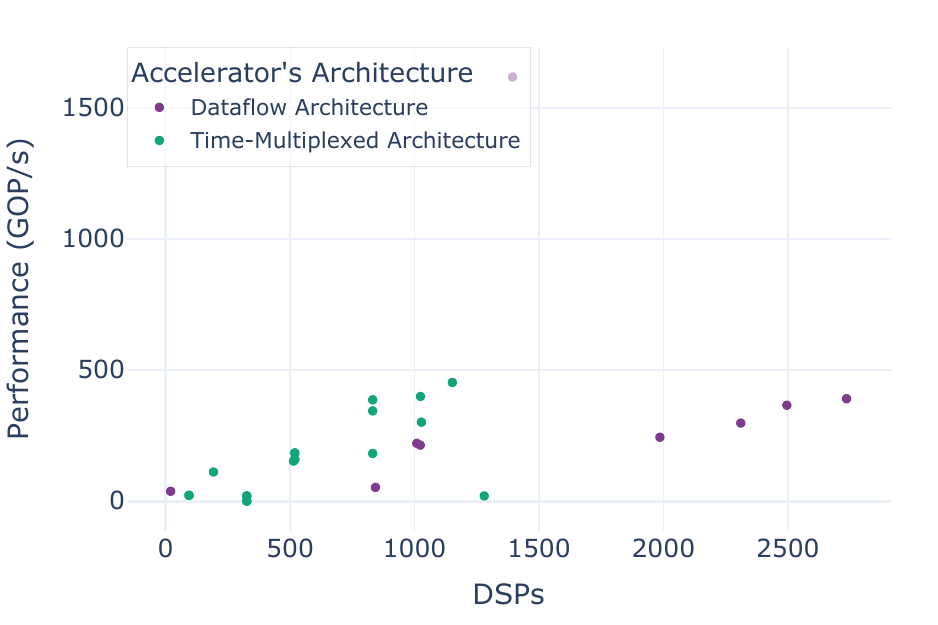}
\caption{DSPs vs. performance (GOP/s)}
\label{fig:dsp_vs_performance}
\end{subfigure}
\hfill
\begin{subfigure}[b]{0.49\textwidth}
\centering
\includegraphics[width=\linewidth]{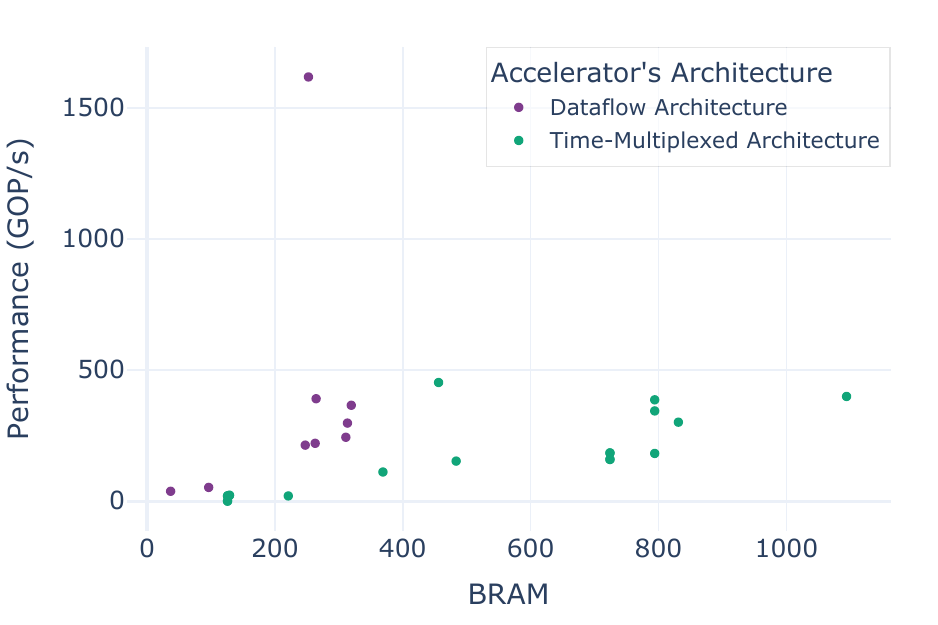}
\caption{BRAM vs. performance (GOP/s)}
\label{fig:bram_vs_performance}
\end{subfigure}
\caption{Influence of DSPs and BRAM on performance}
\label{fig:dsp_bram_performance}
\Description[Influence of DSPs and BRAM on performance]{Influence of DSPs and BRAM on performance. The first figure shows the number of DSPs vs. performance (GOP/s), while the second figure shows the number of BRAM vs. performance (GOP/s). The first and second figures show an increase in performance with the number of DSPs and BRAMs, respectively.}
\end{figure}

Inference times for the surveyed accelerators ranged from under 10 ms to over 100 s per image. Because many lacked pipelining or batch-processing capabilities, their throughput was inversely proportional to inference time. These metrics depend on an NN model's number of operations, parameters, and input size (see Figure~\ref{fig:operations_parameters}). By contrast, OP/s abstracts from these model-specific characteristics and, as shown in Figure~\ref{fig:dsp_bram_performance}, depends primarily on the accelerator's logic design and the FPGA resources it uses. The outlier in this figure leverages FINN to create a highly specific accelerator, resulting in the most efficient implementation in terms of OP/s/W~\cite{liFPGABasedLowBitLightweight2025}. While FPS remains critical for mission planning and model-specific evaluation, OP/s better reflects an accelerator's overall hardware capability, making it a more suitable metric for comparing accelerators across diverse applications. Furthermore, Figure~\ref{fig:bram_vs_performance} shows that dataflow architectures generally consume fewer BRAM resources than time-multiplexed designs.

\begin{wrapfigure}{r}{0.49\textwidth}
\centering
\includegraphics[width=\linewidth]{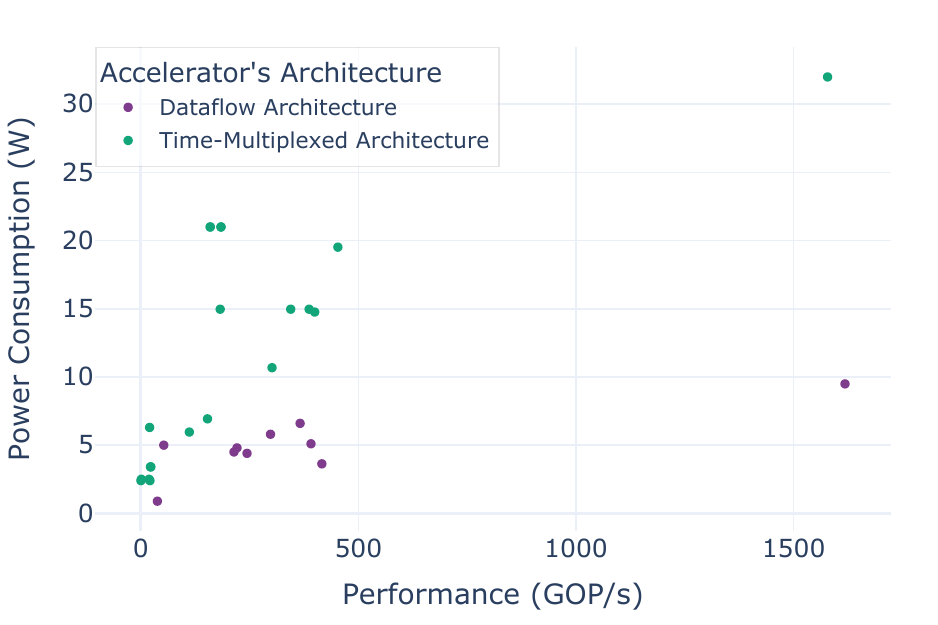}
\caption{\label{fig:performance_vs_power}Performance vs. power consumption}
\Description[Performance vs. power consumption]{Performance vs. power consumption. The figure shows the performance (GOP/s) vs. power consumption (W). We see a general proportionality between performance and power consumption, with some exceptions.}
\end{wrapfigure}

Figure~\ref{fig:performance_vs_power} illustrates a general proportionality between OP/s and accelerator power consumption. This relationship is expected, as higher OP/s typically increase FPGA resource utilization, thereby raising power consumption. The two outliers in this figure demonstrate the scalability and trade-offs of both architectures. To achieve comparable performance, the time-multiplexed architecture~\cite{liuEfficientAcceleratorDNNBased2019} requires more power and resources than its dataflow counterpart~\cite{liFPGABasedLowBitLightweight2025}. Notably, the former accelerates a 27.4 GOP NN, whereas the latter processes a 0.44 GOP NN.

Furthermore, the power values reported in the surveyed literature are often estimated rather than measured, and focus primarily on the chip or accelerator rather than the entire board. Given the criticality of this metric for space applications, we recommend explicitly reporting measured power consumption. When this is unfeasible, studies should employ a robust estimation tool and detail its use in their methodology~\cite{rtl_power_estimation}. Accurate power reporting enables the derivation of OP/s/W. This metric is particularly valuable for on-board spacecraft implementations, as it clearly delineates the relationship between accelerator performance and power efficiency.

\begin{figure}[htbp]
\centering
\begin{subfigure}[b]{0.49\textwidth}
\centering
\includegraphics[width=\linewidth]{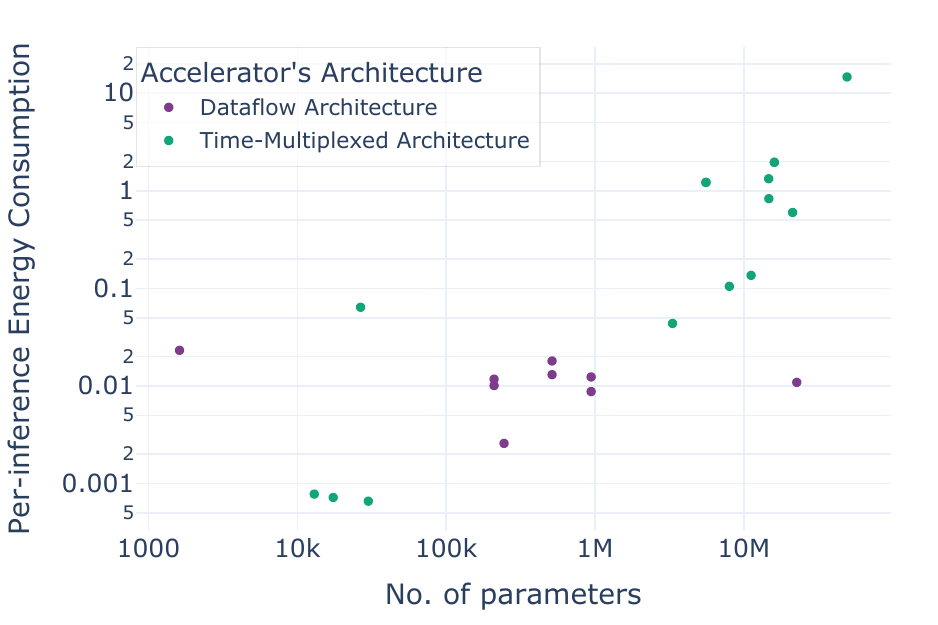}
\caption{Number of parameters vs. per-inference energy consumption}
\label{fig:parameters_vs_energy}
\end{subfigure}
\hfill
\begin{subfigure}[b]{0.49\textwidth}
\centering
\includegraphics[width=\linewidth]{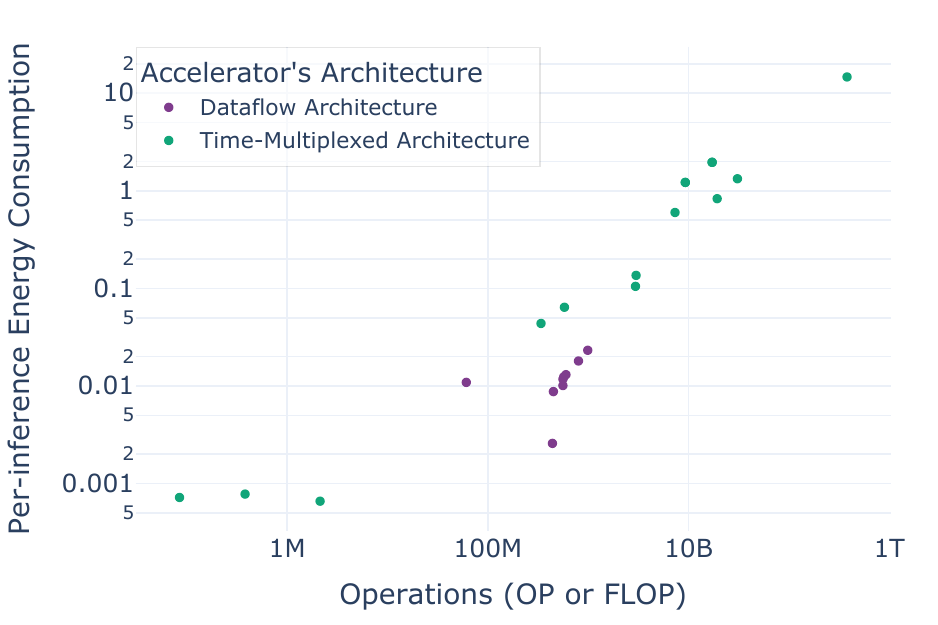}
\caption{Number of operations vs. per-inference energy consumption}
\label{fig:operations_vs_energy}
\end{subfigure}
\caption{Influence of parameters and operations on per-inference energy consumption}
\label{fig:parameters_operations_energy}
\Description[Influence of parameters and operations on per-inference energy consumption]{Influence of parameters and operations on per-inference energy consumption. The first figure shows the number of parameters vs. per-inference energy consumption, while the second figure shows the number of operations vs. per-inference energy consumption. The first figure shows an increase in per-inference energy consumption with the number of parameters. In contrast, the second figure shows an increase in per-inference energy consumption with the number of operations.}
\end{figure}

Power consumption alone does not determine total energy use. An accelerator that consumes more power but executes faster inference might use less energy than one with lower power consumption but longer inference times. While power consumption depends solely on the accelerator architecture, energy consumption also depends on the NN model. Figures~\ref{fig:parameters_vs_energy} and~\ref{fig:operations_vs_energy} show this dependency and indicate that the number of operations influences energy consumption more than the number of parameters. Consequently, deeper NNs with more operations require more energy for inference than those with an equivalent parameter count but fewer operations.

Only 4 surveyed studies employed SNNs~\cite{lemaireFPGAbasedHybridNeural2020,lentEvaluatingCognitiveNetwork2020,abderrahmaneSPLEATSPikingLowpower2022,jiangRFFingerprintingIdentification2023}. These accelerators are omitted from Figures~\ref{fig:operations_parameters}--\ref{fig:parameters_operations_energy} because they do not report the operation counts or GOP/s that those comparisons require. Although SNNs can be more energy-efficient than ANNs under high spike sparsity~\cite{snn_vs_ann}, the surveyed space implementations do not confirm this advantage: converting conventional sensor data into spikes increases latency and therefore per-inference energy.

\subsection{Radiation Concerns, Reliability, and Robustness}
Lastly, 15 of the 47 reviewed papers discussed radiation concerns~\cite{cosmasUtilizationFPGAOnboard2020,rapuanoFPGABasedHardwareAccelerator2021,yangAlgorithmHardwareCodesign2022,nerisFPGABasedImplementationCNN2022,niAlgorithmHardwareCoOptimizationDeployment2023,guoOverlayAcceleratorDeepLab2024,wangHardwareAccelerationImplementation2022,kesumaArtificialIntelligenceImplementation2019,yanAutomaticDeploymentConvolutional2022,sabogalReCoNReconfigurableCNN2019,sabogalReconfigurableFrameworkResilient2021,reiterFPGAAccelerationQuantized2020,gankidiFPGAArchitectureDeep2017,gpu_at_sat,rnn_pacini}, and only a few incorporated mechanisms against SEEs~\cite{sabogalReCoNReconfigurableCNN2019,sabogalReconfigurableFrameworkResilient2021,rapuanoFPGABasedHardwareAccelerator2021,kesumaArtificialIntelligenceImplementation2019,rnn_pacini}. Most of the surveyed accelerators targeted COTS SRAM FPGAs, for which configuration-memory SEUs are the dominant failure mode (Section~\ref{sec:fault_tolerance}). Implementing FPGA-aware fault tolerance---in particular, pairing spatial redundancy, such as TMR, with frequent configuration scrubbing---can affect inference time, throughput, resource usage, and power consumption, underscoring a significant gap in current research. Algorithmic NN techniques (for example, fault-aware training) address datapath errors but do not restore corrupted routing or LUT configuration. Although certain missions may not require additional precautions and can neglect these vulnerabilities, others strictly necessitate radiation-tolerant hardware. Among the reviewed literature, only Sabogal et al.~\cite{sabogalReCoNReconfigurableCNN2019,sabogalReconfigurableFrameworkResilient2021} combined TMR with configuration scrubbing, conducted beam experiments, and reported the failure rate of their accelerator. Nonetheless, separate studies have evaluated the resilience of NN accelerators to SEEs~\cite{ANN_reliability_Space,cnn_see_evaluation}. Furthermore, robustness considerations were absent from the surveyed works. While some missions are not safety-critical~\cite{fogaCloudDetectionAlgorithm2017,giuffridaCloudScoutDeepNeural2020}, others---such as satellite component detection and autonomous navigation~\cite{gankidiFPGAArchitectureDeep2017,cosmasUtilizationFPGAOnboard2020}---may require rigorous methods to certify that their NN outputs guarantee correct behavior. For these mission-critical scenarios, it is important to investigate how to achieve fault reliability and robustness through testing and verification, a practice well established in other domains~\cite{nn_robustness,nn_robustness_automative,nn_verification_survey}.

The limited attention given to radiation effects in most studies underscores the necessity for research on radiation-tolerant NN accelerators. Future work should treat SRAM-FPGA fault tolerance as a distinct problem from CPU and ASIC reliability, document the performance trade-offs of TMR with configuration scrubbing, and report failure rates under fault injection or beam testing. Targeting a radiation-tolerant device is complementary rather than sufficient~\cite{rnn_pacini}: flash-based families such as PolarFire~\cite{PolarFire} largely avoid configuration SEUs and can rely on ASIC-like protection of user state, whereas SRAM families---including COTS UltraScale+ devices and radiation-tolerant Kintex UltraScale FPGAs~\cite{xilinx2020rt}---still require configuration repair for high reliability~\cite{fpga_reliability}.

\section{Conclusion}
\label{sec:Section_5}
This survey provides a comprehensive overview of FPGA-based NN accelerators for on-board spacecraft computing, based on an analysis of 47 recent publications. It highlights increasing work on FPGA-based NN acceleration for space use cases. Most of these accelerators still implement CNN architectures, particularly classification models for EO. Most studies utilized AMD FPGAs and adopted either time-multiplexed or dataflow architectures, demonstrating diverse design approaches. Performance evaluation consistently emphasized inference time, throughput, and power consumption, reflecting the critical importance of energy efficiency in space missions. The survey's 2024 cutoff coincides with a shift in the broader AI community from CNNs toward Transformer architectures, large language models, and generative methods~\cite{transformers,gan_rs_survey,diffusion_rs}, which future FPGA accelerators for space might need to support. The current state of research underscores significant opportunities for future exploration, including broader NN architectures, improved on-board training methods, and enhanced radiation-tolerance mechanisms. Addressing these aspects will be crucial to advancing and deploying FPGA-based NN accelerators in next-generation space missions.

\begin{acks}
  We thank the anonymous reviewers for their valuable feedback. This work is supported by the European Commission, with Automatics in Space Exploration (ASAP), project no. 101082633. We used generative AI to draft and edit selected passages. AI was \textbf{not} used to summarize the literature. The authors reviewed and verified all technical content, results, and citations.
\end{acks}

\bibliographystyle{ACM-Reference-Format}
\bibliography{bibliography,other}

\newpage
\appendix
\section{Algorithm and Hardware metrics extracted from the reviewed literature}
\label{tab:evaluation_metrics}
\begin{table*}[htbp]
    \centering
    \renewcommand{\arraystretch}{1.3} 
\begin{tabular}{@{} p{0.48\textwidth} p{0.48\textwidth} @{}}
        \toprule
        \textbf{Algorithm-related Metrics} & \textbf{Hardware-related Metrics} \\
        \midrule
        
        \textbf{Application:} Identifies the specific space-re\-lated task. & 
        \textbf{Board:} Specifies the target development platform. \\
        
        \textbf{Task Type/s:} Learning objective the model is designed to solve. & 
        \textbf{FPGA:} Specifies the exact silicon chip utilized. \\
        
        \textbf{NN Architecture:} The specific underlying neural network structure (e.g., CNN, SNN, MLP). & 
        \textbf{Hardware Implementation:} Details the hardware description language or tool used. \\
        
        \textbf{Data Origin:} Identifies the source of the data processed. & 
        \textbf{Accelerator Architecture (NN Flexibility):} Characterizes the main hardware design paradigm (e.g., dataflow or time-multiplexed). (Defines if the accelerator can run different NN models, a specific model with variable parameters, or a specific model with specific parameters, without changing the FPGA design.) \\
        
        \textbf{Dataset/s:} Details the actual data compilation used for training and benchmarking. & 
        \textbf{FPGA Resource Consumption:} Quantifies the utilization of on-chip components, including LUTs, FFs, DSPs, and BRAMs. \\
        
        \textbf{Training Method:} Specifies whether training is on-board or off-board, and if supervised or unsupervised. & 
        \textbf{Frequency (MHz):} Tracks the operating clock speed of the synthesized hardware design. \\
        
        \textbf{Input Size:} Defines the dimensions of the images or data arrays fed into the network. & 
        \textbf{Inference Time (Latency):} Measures the absolute time required to process a single input. \\
\bottomrule
    \end{tabular}
\end{table*}
\begin{landscape}
\section{Algorithm-related table}\label{app:algorithm_table}
\begin{center}
\scalebox{0.6}{
\renewcommand{\arraystretch}{1.3}
\begin{tabular}{lllllllllllllll}
\toprule
\textbf{Key.} & \textbf{Year} & \textbf{Application} & \makecell[tl]{\textbf{Task}\\\textbf{Type/s}} & \makecell[tl]{\textbf{NN}\\\textbf{Architecture}} & \makecell[tl]{\textbf{Data}\\\textbf{Origin}} & \textbf{Dataset/s} & \makecell[tl]{\textbf{Training}\\\textbf{Method}} & \makecell[tl]{\textbf{Input}\\\textbf{Size}} & \makecell[tl]{\textbf{No. of}\\\textbf{Parameters}} & \makecell[tl]{\textbf{Arithmetic}\\\textbf{Precision}} & \makecell[tl]{\textbf{Memory}\\\textbf{Size}} & \makecell[tl]{\textbf{\#Operations}\\\textbf{(OP or FLOP)}} & \makecell[tl]{\textbf{Quantization}\\\textbf{Method}} & \makecell[tl]{\textbf{Classification}} \\ 
\midrule
\cite{zhuangRealTimeRunway2014} & 2014 & \makecell[tl]{Detect airport\\ runways from\\ EO imagery} & Segmentation & (M)PCNN & Remote & \makecell[tl]{Terrain data\\ containing RGB\\ images of\\ runways} & \makecell[tl]{Off-board\\ and\\ Supervised} & \makecell[tl]{400x400x3\\ (RGB Image)} & --- & --- & --- & --- & --- & \makecell[tl]{TP Accuracy 88\%\\ TN Accuracy 90\%}  \\ \hline
\cite{gankidiFPGAArchitectureDeep2017} & 2017 & \makecell[tl]{Onboard spacecraft\\ autonomy and\\ decision-making} & Classification & MLP & In-Situ & --- & \makecell[tl]{On-board\\ with\\ Q-learning} & \makecell[tl]{11\\ ---1---\\ 25} & --- & \makecell[tl]{Floating-point\\ ---2---\\ Fixed-point} & --- & --- & --- & --- \\ \hline
\cite{chenFPGABasedImplementation2018} & 2018 & \makecell[tl]{Classify pixels\\ in a hyperspectral\\ image} & Classification & CNN & Remote & \makecell[tl]{Pavia University\\ Dataset} & \makecell[tl]{Off-board\\ and\\ Supervised} & \makecell[tl]{1 pixel\\ (610x340\\ images)} & 187,240 & FP32 & --- & --- & --- & \makecell[tl]{Average Acc.\\ 94.58\%\\ Overall Acc.\\ 92.24\%}  \\
\bottomrule
\end{tabular}}
\end{center}

\begin{center}
\scalebox{0.6}{
\renewcommand{\arraystretch}{1.3}
\begin{tabular}{lllllllllllllll}
\toprule
\textbf{Key.} & \textbf{Year} & \textbf{Application} & \makecell[tl]{\textbf{Task}\\\textbf{Type/s}} & \makecell[tl]{\textbf{NN}\\\textbf{Architecture}} & \makecell[tl]{\textbf{Data}\\\textbf{Origin}} & \textbf{Dataset/s} & \makecell[tl]{\textbf{Training}\\\textbf{Method}} & \makecell[tl]{\textbf{Input}\\\textbf{Size}} & \makecell[tl]{\textbf{No. of}\\\textbf{Parameters}} & \makecell[tl]{\textbf{Arithmetic}\\\textbf{Precision}} & \makecell[tl]{\textbf{Memory}\\\textbf{Size}} & \makecell[tl]{\textbf{\#Operations}\\\textbf{(OP or FLOP)}} & \makecell[tl]{\textbf{Quantization}\\\textbf{Method}} & \makecell[tl]{\textbf{Classification}} \\ 
\midrule
\cite{bahlLowpowerNeuralNetworks2019} & 2019 & \makecell[tl]{Semantic\\ segmentation on\\ low-power devices} & Segmentation & \makecell[tl]{FCN (CNN\\ without fully\\ connected layers)} & Remote & \makecell[tl]{Cloud-38,\\ CloudPeru2} & \makecell[tl]{Off-board\\ and\\ Supervised} & \makecell[tl]{2048x1944x3\\ (RGB Image)} & 273 & FP32 & 47 KB & 4,654 FLOP & --- & \makecell[tl]{Acc.\\ 93.23, 91.37\%\\ Prec. \\91.67, 93.50\%\\ Rec. \\85.45, 87.76\%}  \\ \hline
\cite{reiterFPGAAccelerationQuantized2020} & 2020 & \makecell[tl]{Real-time cloud\\ detection on EO\\ images} & Classification & CNN & Remote & \makecell[tl]{Custom ESA\\ Copernicus\\ Sentinel-2 \\ and the “Planet:\\ Understanding\\ the Amazon from\\ Space” datasets} & \makecell[tl]{Off-board\\ and\\ Supervised} & \makecell[tl]{32x32x3\\ (RGB image)} & --- & 1-bit & --- & --- & QAT & Acc. 64 \% \\ \hline
\cite{lemaireFPGAbasedHybridNeural2020} & 2020 & \makecell[tl]{Cloud detection\\ onboard LEO\\ satellites} & Classification & CNN + SNN & Remote & \makecell[tl]{Custom\\ satellite\\ image dataset} & \makecell[tl]{Off-board\\ and\\ Supervised} & \makecell[tl]{28x28x3\\ (RGB image)} & --- & \makecell[tl]{16-bit\\ fixed-point} & --- & --- & --- & Acc. 87 \% \\
\bottomrule
\end{tabular}}
\end{center}

\begin{center}
\scalebox{0.6}{
\renewcommand{\arraystretch}{1.3}
\begin{tabular}{lllllllllllllll}
\toprule
\textbf{Key.} & \textbf{Year} & \textbf{Application} & \makecell[tl]{\textbf{Task}\\\textbf{Type/s}} & \makecell[tl]{\textbf{NN}\\\textbf{Architecture}} & \makecell[tl]{\textbf{Data}\\\textbf{Origin}} & \textbf{Dataset/s} & \makecell[tl]{\textbf{Training}\\\textbf{Method}} & \makecell[tl]{\textbf{Input}\\\textbf{Size}} & \makecell[tl]{\textbf{No. of}\\\textbf{Parameters}} & \makecell[tl]{\textbf{Arithmetic}\\\textbf{Precision}} & \makecell[tl]{\textbf{Memory}\\\textbf{Size}} & \makecell[tl]{\textbf{\#Operations}\\\textbf{(OP or FLOP)}} & \makecell[tl]{\textbf{Quantization}\\\textbf{Method}} & \makecell[tl]{\textbf{Classification}} \\ 
\midrule
\cite{kyriakosResourcesPowerEfficient2022} & 2022 & \makecell[tl]{Detect ships\\ presence in\\ a given image} & Classification & CNN & Remote & \makecell[tl]{Planet's ``Ships\\ in Satellite\\ Imagery'' dataset} & \makecell[tl]{Off-board\\ and\\ Supervised} & \makecell[tl]{80x80x3\\ (RGB image)} & 84,000 & \makecell[tl]{8 to 17-bit\\ fixed-point} & --- & 18,122,000 OP & PTQ & Acc. 97.6 \% \\ \hline
\cite{nerisFPGABasedImplementationCNN2022} & 2022 & \makecell[tl]{Detect ships\\ and airplanes} & \makecell[tl]{Object\\ Detection} & \makecell[tl]{(CNN)\\ MobileNetV1Lite} & Remote & \makecell[tl]{Ship datasets\\ ---1---\\ Airplane datasets} & \makecell[tl]{Off-board\\ and\\ Supervised} & \makecell[tl]{512x512x3\\ ---1---\\ 256x256x3\\ (RGB image)} & 120,000 & \makecell[tl]{FP32\\ ----2----\\ 16-bit\\ fixed-point} & --- & \makecell[tl]{50 MFLOP\\ ---1---\\ 12 MFLOP} & PTQ & \makecell[tl]{$\text{F1}_{\text{score}} \approx 94.5\%$\\ ---1---\\ $F1_{Score}\approx 91\%$}  \\
\bottomrule
\end{tabular}}
\end{center}

\begin{center}
\scalebox{0.6}{
\renewcommand{\arraystretch}{1.3}
\begin{tabular}{lllllllllllllll}
\toprule
\textbf{Key.} & \textbf{Year} & \textbf{Application} & \makecell[tl]{\textbf{Task}\\\textbf{Type/s}} & \makecell[tl]{\textbf{NN}\\\textbf{Architecture}} & \makecell[tl]{\textbf{Data}\\\textbf{Origin}} & \textbf{Dataset/s} & \makecell[tl]{\textbf{Training}\\\textbf{Method}} & \makecell[tl]{\textbf{Input}\\\textbf{Size}} & \makecell[tl]{\textbf{No. of}\\\textbf{Parameters}} & \makecell[tl]{\textbf{Arithmetic}\\\textbf{Precision}} & \makecell[tl]{\textbf{Memory}\\\textbf{Size}} & \makecell[tl]{\textbf{\#Operations}\\\textbf{(OP or FLOP)}} & \makecell[tl]{\textbf{Quantization}\\\textbf{Method}} & \makecell[tl]{\textbf{Classification}} \\ 
\midrule
\cite{ekbladResourceconstrainedFPGADesign2023} & 2023 & \makecell[tl]{Satellite\\ component\\ detection} & \makecell[tl]{Object\\ Detection} & \makecell[tl]{(CNN)\\ modified YOLOv4} & Remote & \makecell[tl]{Custom\\ dataset} & \makecell[tl]{Off-board\\ and\\ Supervised} & \makecell[tl]{416x416x3\\ (RGB image)} & --- & \makecell[tl]{8-bit\\ fixed point} & --- & --- & --- & \makecell[tl]{mAP\\ @(IoU 0.5)\\ 6.3\%} \\ \hline
\cite{carmeliAIBasedRealTimeStar2023} & 2023 & \makecell[tl]{Identifying\\ star patterns\\ from images\\ captured by\\ onboard satellite\\ cameras} & Classification & SOM & Remote & \makecell[tl]{Custom\\ dataset} & \makecell[tl]{Off-board\\ and\\ Supervised} & 10 & --- & --- & 249 KB & --- & --- & Acc. 98\% \\ \hline
\cite{cocaFPGAAcceleratorMetaRecognition2023} & 2023 & \makecell[tl]{LEO\\(Anomaly\\ detection)} & Classification & \makecell[tl]{(CNN)\\ ResNet-50} & Remote & \makecell[tl]{BigEarthNet,\\ Sentinel-2\\ images (Zamora\\ and Bordeaux)} & \makecell[tl]{Off-board\\ and\\ Supervised} & \makecell[tl]{120x120x10\\ (multispectral\\ image)} & --- & 8-bit & --- & --- & Finetuned & \makecell[tl]{Precision =\\ 73.7, 72.7,\\ 78.7\%} \\
\bottomrule
\end{tabular}}
\end{center}

\begin{center}
\scalebox{0.6}{
\renewcommand{\arraystretch}{1.3}
\begin{tabular}{lllllllllllllll}
\toprule
\textbf{Key.} & \textbf{Year} & \textbf{Application} & \makecell[tl]{\textbf{Task}\\\textbf{Type/s}} & \makecell[tl]{\textbf{NN}\\\textbf{Architecture}} & \makecell[tl]{\textbf{Data}\\\textbf{Origin}} & \textbf{Dataset/s} & \makecell[tl]{\textbf{Training}\\\textbf{Method}} & \makecell[tl]{\textbf{Input}\\\textbf{Size}} & \makecell[tl]{\textbf{No. of}\\\textbf{Parameters}} & \makecell[tl]{\textbf{Arithmetic}\\\textbf{Precision}} & \makecell[tl]{\textbf{Memory}\\\textbf{Size}} & \makecell[tl]{\textbf{\#Operations}\\\textbf{(OP or FLOP)}} & \makecell[tl]{\textbf{Quantization}\\\textbf{Method}} & \makecell[tl]{\textbf{Classification}} \\ 
\midrule
\cite{shaoConfigurableAcceleratorCNNBased2024} & 2024 & \makecell[tl]{Object Detection\\ Onboard Satellites} & \makecell[tl]{Object\\ Detection} & \makecell[tl]{(CNN)\\ ResNet-18\\ ---------\\ YOLOv3-Tiny} & Remote & NWPU VHR-10 & \makecell[tl]{Off-board\\ and\\ Supervised} & \makecell[tl]{256x256x3\\ (RGB Image)} & \makecell[tl]{11,200,000\\ ---------\\ 8,000,000} & \makecell[tl]{16-bit\\ ---------\\ 8-bit} & --- & \makecell[tl]{* 3 GOP\\ ---------\\ * 2.96 GOP} & --- & \makecell[tl]{Acc. 81 \%\\ ---------\\ Acc. 96 \%}  \\ \hline
\cite{kimFPGAAcceleratedCNNParallelized2024} & 2024 & \makecell[tl]{Routing in a\\ LEO satellite\\ network} & Classification & CNN & In-Situ & \makecell[tl]{Custom\\ Dataset} & \makecell[tl]{Off-board\\ and\\ Supervised} & 4x15x15 & --- & --- & --- & --- & --- & --- \\
\bottomrule
\end{tabular}}
\end{center}

\section{Hardware-related table}\label{app:hardware_table}
\begin{center}
\scalebox{0.6}{
\renewcommand{\arraystretch}{1.3}
\begin{tabular}{llllllllllllll}
\toprule
\textbf{Key.} & \textbf{Board} & \textbf{FPGA} & \makecell[tl]{\textbf{Accelerator's}\\ \textbf{Architecture}\\ \textbf{(NN Flexibility)}} & \makecell[tl]{\textbf{Hardware}\\ \textbf{Implementation}} & \makecell[tl]{\textbf{FPGA's}\\ \textbf{Resource}\\ \textbf{Consumption}} & \makecell[tl]{\textbf{Frequency}} & \makecell[tl]{\textbf{Inference}\\ \textbf{Time}\\ \textbf{(Latency)}} & \textbf{Throughput} & \makecell[tl]{\textbf{Performance}} & \makecell[tl]{\textbf{Power}\\ \textbf{Consumption}} & \makecell[tl]{\textbf{Power}\\ \textbf{Efficiency}} & \makecell[tl]{\textbf{Per-inference}\\ \textbf{Energy}\\ \textbf{Consumption}} & \makecell[tl]{\textbf{Radiation}\\ \textbf{Tolerance}}  \\ \midrule
\cite{zhuangRealTimeRunway2014} & --- & \makecell[tl]{Spartan-3A\\ XC3SD3400A\\-4FGG676C} & \makecell[tl]{---\\ (Fixed)} & --- & \makecell[tl]{2,958,000\\ Logic Cells} & 250 MHz & \makecell[tl]{15.6 ms\\ ---------\\ 31.1 ms} & --- & --- & --- & --- & --- & --- \\ \hline
\cite{gankidiFPGAArchitectureDeep2017} & --- & Virtex 7 & \makecell[tl]{Dataflow\\ Architecture\\ (Fixed)} & --- & --- & 150 MHz & \makecell[tl]{0.9 $\mu s$\\ ---2---\\ 13 $\mu s$\\ ---1---\\ 4 $\mu s$\\ ---2---\\ 107 $\mu s$} & --- & --- & \makecell[tl]{5.6 W\\ ---2---\\ 7.1 W\\ ---1---\\ 7.1 W\\ ---2---\\ 9.7 W} & --- & \makecell[tl]{* 5.04 $\mu J$\\ ---2---\\ * 92.3 $\mu J$\\ ---1---\\ * 28.4 $\mu J$\\ ---2---\\ * 1037.9 $\mu J$} & \makecell[tl]{RT-\\Comparable} \\ \hline
\cite{chenFPGABasedImplementation2018} & --- & --- & \makecell[tl]{Dataflow\\ Architecture\\ (Fixed)} & Unknown HDL & \makecell[tl]{616,660 LUT\\ 1,286,528 FF\\ 15,260 DSP\\ 64.47 kB MRAM} & 200 MHz & 207.607 s & 0.0048 FPS & --- & --- & --- & --- & --- \\
\bottomrule
\end{tabular}}
\end{center}

\begin{center}
\scalebox{0.6}{
\renewcommand{\arraystretch}{1.3}
\begin{tabular}{llllllllllllll}
\toprule
\textbf{Key.} & \textbf{Board} & \textbf{FPGA} & \makecell[tl]{\textbf{Accelerator's}\\ \textbf{Architecture}\\ \textbf{(NN Flexibility)}} & \makecell[tl]{\textbf{Hardware}\\ \textbf{Implementation}} & \makecell[tl]{\textbf{FPGA's}\\ \textbf{Resource}\\ \textbf{Consumption}} & \makecell[tl]{\textbf{Frequency}} & \makecell[tl]{\textbf{Inference}\\ \textbf{Time}\\ \textbf{(Latency)}} & \textbf{Throughput} & \makecell[tl]{\textbf{Performance}} & \makecell[tl]{\textbf{Power}\\ \textbf{Consumption}} & \makecell[tl]{\textbf{Power}\\ \textbf{Efficiency}} & \makecell[tl]{\textbf{Per-inference}\\ \textbf{Energy}\\ \textbf{Consumption}} & \makecell[tl]{\textbf{Radiation}\\ \textbf{Tolerance}}  \\ \midrule
\cite{liuEfficientAcceleratorDNNBased2019} & \makecell[tl]{Intel's\\ Arria 10\\ SOC} & A10-SX 660 & \makecell[tl]{Time-Multiplexed\\ Architecture\\ (Weight Flexible)} & \makecell[tl]{Verilog\\ HDL} & \makecell[tl]{170,906 LUT\\ 377,142 FF\\ 1,665 DSP\\ 1,894 M20K} & 200 MHz & 17.4 ms & 57 FPS & 1,578 GOP/s & \makecell[tl]{32 W\\ (the board)} & \makecell[tl]{49.3\\ GOP/s/W} & * 556.8 mJ & --- \\ \hline
\cite{liEfficientObjectDetection2019} & \makecell[tl]{Zynq 7000\\ Board} & \makecell[tl]{XC7Z100\\ FPGA} & \makecell[tl]{Time-Multiplexed\\ Architecture\\ (Weight Flexible)} & \makecell[tl]{Verilog\\ HDL} & \makecell[tl]{136,892 LUT\\ 156,238 FF\\ 1,152 DSP\\ 912 BRAM\_18k} & 200 MHz & 42.59 ms & * 23 FPS & 452.8 GOP/s & 19.52 W & \makecell[tl]{23.20\\ GOP/s/W} & * 831 mJ & --- \\ \hline
\cite{bahlLowpowerNeuralNetworks2019} & --- & \makecell[tl]{Cyclone V\\ 5CSXC6\\ FPGA} & \makecell[tl]{Dataflow\\ Architecture\\ (Fixed)} & VGT & --- & 100 MHz & < 150 ms & > 7 FPS & --- & --- & --- & --- & --- \\
\bottomrule
\end{tabular}}
\end{center}

\begin{center}
\scalebox{0.6}{
\renewcommand{\arraystretch}{1.3}
\begin{tabular}{llllllllllllll}
\toprule
\textbf{Key.} & \textbf{Board} & \textbf{FPGA} & \makecell[tl]{\textbf{Accelerator's}\\ \textbf{Architecture}\\ \textbf{(NN Flexibility)}} & \makecell[tl]{\textbf{Hardware}\\ \textbf{Implementation}} & \makecell[tl]{\textbf{FPGA's}\\ \textbf{Resource}\\ \textbf{Consumption}} & \makecell[tl]{\textbf{Frequency}} & \makecell[tl]{\textbf{Inference}\\ \textbf{Time}\\ \textbf{(Latency)}} & \textbf{Throughput} & \makecell[tl]{\textbf{Performance}} & \makecell[tl]{\textbf{Power}\\ \textbf{Consumption}} & \makecell[tl]{\textbf{Power}\\ \textbf{Efficiency}} & \makecell[tl]{\textbf{Per-inference}\\ \textbf{Energy}\\ \textbf{Consumption}} & \makecell[tl]{\textbf{Radiation}\\ \textbf{Tolerance}}  \\ \midrule
\cite{paciniMultiCacheSystemOnChip2021} & \makecell[tl]{Zynq\\ UltraScale+\\ MPSoC\\ ZCU106} & XCZU7EV & \makecell[tl]{Time-Multiplexed\\ Architecture\\ (Weight Flexible)} & VHDL & \makecell[tl]{59,195 LUT\\ 1128 LUTRAM\\ 20,771 FF\\ 1158 DSP\\ 39 BRAM\\ 18 URAM} & \makecell[tl]{$115.4_{acc}$,\\ $200_{axi}$\\ MHz} & 144.8 ms & * 6.91 FPS & --- & 4.51 W & --- & 0.65 J & --- \\
\bottomrule
\end{tabular}}
\end{center}

\begin{center}
\scalebox{0.6}{
\renewcommand{\arraystretch}{1.3}
\begin{tabular}{llllllllllllll}
\toprule
\textbf{Key.} & \textbf{Board} & \textbf{FPGA} & \makecell[tl]{\textbf{Accelerator's}\\ \textbf{Architecture}\\ \textbf{(NN Flexibility)}} & \makecell[tl]{\textbf{Hardware}\\ \textbf{Implementation}} & \makecell[tl]{\textbf{FPGA's}\\ \textbf{Resource}\\ \textbf{Consumption}} & \makecell[tl]{\textbf{Frequency}} & \makecell[tl]{\textbf{Inference}\\ \textbf{Time}\\ \textbf{(Latency)}} & \textbf{Throughput} & \makecell[tl]{\textbf{Performance}} & \makecell[tl]{\textbf{Power}\\ \textbf{Consumption}} & \makecell[tl]{\textbf{Power}\\ \textbf{Efficiency}} & \makecell[tl]{\textbf{Per-inference}\\ \textbf{Energy}\\ \textbf{Consumption}} & \makecell[tl]{\textbf{Radiation}\\ \textbf{Tolerance}}  \\ \midrule
\cite{pitonakCloudSatNet1FPGABasedHardwareAccelerated2022} & Zturn & \makecell[tl]{Zynq\\ Z7020} & \makecell[tl]{Dataflow\\ Architecture\\ (Fixed)} & FINN & \makecell[tl]{25,387 LUT\\ 34,569 FF\\ 30 DSP\\ 46 BRAM\\ ---------\\ 36,735 LUT\\ 51,721 FF\\ 30 DSP\\ 87.5 BRAM} & 100 MHz & --- & \makecell[tl]{15.462 FPS\\ ---------\\ 15.468 FPS} & --- & \makecell[tl]{2.520 W\\ ---------\\ 2.592 W} & --- & --- & --- \\ \hline
\cite{zhangAccurateLowlatencyEfficient2022} & \makecell[tl]{Zynq\\ UltraScale+\\ MPSoC\\ ZCU104} & ZU7EV & \makecell[tl]{Time-Multiplexed\\ Architecture\\ (Fixed)} & --- & \makecell[tl]{178,000 LUT\\ 1,280 DSP\\ 221 BRAM\\ 96 URAM} & 125 MHz & 0.105 ms & 9500 FPS & * 20.24 GOP/s & 6.3 W & \makecell[tl]{* 3.21\\ GOP/s/W} & 0.66 mJ & --- \\ \hline
\cite{abderrahmaneSPLEATSPikingLowpower2022} & \makecell[tl]{* MitySOM-\\-5CSX} & Cyclone V & \makecell[tl]{Dataflow\\ Architecture\\ (Fixed)} & VHDL & --- & --- & \makecell[tl]{0.025 ms\\ ---------\\  0.28 ms\\ ---------\\ 0.16 - 1.86 ms} & --- & --- & \makecell[tl]{2.90 W\\ ---------\\  2.96 W\\ ---------\\ 2.66 W} & --- & \makecell[tl]{* 0.07 mJ\\ ---------\\ * 0.83 mJ\\ ---------\\ * 0.43 - 4.95 mJ} & --- \\
\bottomrule
\end{tabular}}
\end{center}

\begin{center}
\scalebox{0.6}{
\renewcommand{\arraystretch}{1.3}
\begin{tabular}{llllllllllllll}
\toprule
\textbf{Key.} & \textbf{Board} & \textbf{FPGA} & \makecell[tl]{\textbf{Accelerator's}\\ \textbf{Architecture}\\ \textbf{(NN Flexibility)}} & \makecell[tl]{\textbf{Hardware}\\ \textbf{Implementation}} & \makecell[tl]{\textbf{FPGA's}\\ \textbf{Resource}\\ \textbf{Consumption}} & \makecell[tl]{\textbf{Frequency}} & \makecell[tl]{\textbf{Inference}\\ \textbf{Time}\\ \textbf{(Latency)}} & \textbf{Throughput} & \makecell[tl]{\textbf{Performance}} & \makecell[tl]{\textbf{Power}\\ \textbf{Consumption}} & \makecell[tl]{\textbf{Power}\\ \textbf{Efficiency}} & \makecell[tl]{\textbf{Per-inference}\\ \textbf{Energy}\\ \textbf{Consumption}} & \makecell[tl]{\textbf{Radiation}\\ \textbf{Tolerance}}  \\ \midrule
\cite{zhaoHardwareAccelerationSatellite2023} & \makecell[tl]{Kria KV260\\ Vision AI\\ Starter Kit} & XCK26 & \makecell[tl]{Time-Multiplexed\\ Architecture\\ (Flexible)} & \makecell[tl]{AMD DPU + \\Vitis AI} & \makecell[tl]{53,540 LUT\\ 710 DSP} & 200 MHz & * 21 ms & 48.14 FPS & --- & 7.2 W & --- & * 151 mJ & --- \\ \hline
\cite{niAlgorithmHardwareCoOptimizationDeployment2023} & \makecell[tl]{Virtex 7\\ VC709} & XC7VLX690T & \makecell[tl]{Time-Multiplexed\\ Architecture\\ (Flexible)} & \makecell[tl]{SystemVerilog\\ HDL} & \makecell[tl]{105,509 LUT\\ 282,807 FF\\ 794 BRAM\\ 832 DSP} & 200 MHz & \makecell[tl]{981.4 ms\\ ---1---\\ 89.1 ms\\ ---2---\\ 40.2 ms} & \makecell[tl]{* 1 FPS\\ ---1---\\ * 11 FPS\\ ---2---\\ * 24 FPS} & \makecell[tl]{386.74 GOP/s\\ ---1---\\ 344.44 GOP/s\\ ---2---\\ 182.34 GOP/s} & 14.97 W & \makecell[tl]{25.83 GOP/s/W\\ ---1---\\ 23.01 GOP/s/W\\ ---2---\\ 12.18 GOP/s/W} & \makecell[tl]{* 14.69 J\\ ---1---\\ * 1.33 J\\ ---2---\\ * 0.60 J} & \makecell[tl]{Discussed\\ RT-FPGA} \\
\bottomrule
\end{tabular}}
\end{center}

\begin{center}
\scalebox{0.6}{
\renewcommand{\arraystretch}{1.3}
\begin{tabular}{llllllllllllll}
\toprule
\textbf{Key.} & \textbf{Board} & \textbf{FPGA} & \makecell[tl]{\textbf{Accelerator's}\\ \textbf{Architecture}\\ \textbf{(NN Flexibility)}} & \makecell[tl]{\textbf{Hardware}\\ \textbf{Implementation}} & \makecell[tl]{\textbf{FPGA's}\\ \textbf{Resource}\\ \textbf{Consumption}} & \makecell[tl]{\textbf{Frequency}} & \makecell[tl]{\textbf{Inference}\\ \textbf{Time}\\ \textbf{(Latency)}} & \textbf{Throughput} & \makecell[tl]{\textbf{Performance}} & \makecell[tl]{\textbf{Power}\\ \textbf{Consumption}} & \makecell[tl]{\textbf{Power}\\ \textbf{Efficiency}} & \makecell[tl]{\textbf{Per-inference}\\ \textbf{Energy}\\ \textbf{Consumption}} & \makecell[tl]{\textbf{Radiation}\\ \textbf{Tolerance}}  \\ \midrule
\cite{gpu_at_sat} & \makecell[tl]{Zynq\\ UltraScale+\\ MPSoC\\ ZCU104} & ZU7EV & \makecell[tl]{---\\ (Flexible)} & GPU@SAT & \makecell[tl]{140,721 LUT\\ 10,201 LUTRAM\\ 220,804 FF\\ 312 BRAM\\ 1 URAM\\ 256 DSP} & 250 MHz & --- & 35 FPS & --- & --- & --- & --- & \makecell[tl]{GPU@SAT has\\ been\\ synthesized\\ to an\\ RT-FPGA~\cite{gpu_at_sat_rt_fpga}}  \\
\bottomrule
\end{tabular}}
\end{center}

\end{landscape}

\end{document}